\documentclass{article}

\usepackage{PRIMEarxiv}

\usepackage[utf8]{inputenc} % allow utf-8 input
\usepackage[T1]{fontenc}    % use 8-bit T1 fonts
\usepackage{hyperref}       % hyperlinks
\usepackage{url}            % simple URL typesetting
\usepackage{booktabs}       % professional-quality tables
\usepackage{amsfonts}       % blackboard math symbols
\usepackage{nicefrac}       % compact symbols for 1/2, etc.
\usepackage{microtype}      % microtypography
\usepackage{lipsum}
\usepackage{fancyhdr}       % header
\usepackage{graphicx}       % graphics
\usepackage{natbib}
\usepackage{amsmath}
\graphicspath{{media/}}     % organize your images and other figures under media/ folder

\title{Reconstructing Carbon Monoxide Reanalysis with Machine Learning}

\author{
  Paula Harder, Johannes Flemming\\
  European Center of Medium-Range Weather Forecast (ECMWF), Bonn, Germany\\
}

\begin{document}
\maketitle

%% These dates will be inserted by Copernicus Publications during the typesetting process.

\maketitle

\begin{abstract}
The Copernicus Atmospheric Monitoring Service provides reanalysis products for atmospheric composition by combining model simulations with satellite observations. The quality of these products depends strongly on the availability of the observational data, which can vary over time as new satellite instruments become available or are discontinued, such as Carbon Monoxide (CO) observations of the Measurements Of Pollution In The Troposphere (MOPITT) satellite in early 2025. Machine learning offers a promising approach to compensate for such data losses by learning systematic discrepancies between model configurations. In this study, we investigate machine learning methods to predict monthly-mean total column of Carbon Monoxide re-analysis from  a control model simulation.
\end{abstract}

\section{Introduction}

The Copernicus Atmosphere Monitoring Service (CAMS) global atmospheric composition (AC) reanalysis (EAC4) \citep{Inness2019EAC4} assimilates multiple satellite retrievals of aerosol optical depth, ozone, and nitrogen dioxide, as well as total column carbon monoxide (TCCO) from the MOPITT instrument \citep{Deeter2003MOPITT}, which serves as the only observational CO data source in the system. EAC4 spans the period from 2003 to near present and has been widely used for the analysis of AC anomalies and long-term trends. The termination of MOPITT operations in January 2025 led to a substantial shift in CO fields in the subsequent EAC4 reanalysis, preventing its direct use for diagnosing TCCO anomalies in 2025.

To address this discontinuity, we develop a machine learning-based approach to emulate the impact of MOPITT-driven data assimilation in EAC4. The ML model predicts monthly mean total column carbon monoxide (TCCO) fields by learning the relationship between the EAC4 reanalysis and a corresponding control simulation without data assimilation. The control simulation uses the same meteorological forcing and emissions as EAC4, including wildfire CO emissions, which dominate interannual variability.

%We compare different ML approaches such as linear regression, random forest, gradient boosting and neural networks. 
Recent studies have demonstrated the potential of machine learning to correct systematic biases in atmospheric and chemistry–climate model outputs by learning nonlinear relationships between model simulations and observational or reanalysis data \citep{lary2009ml_bias_modis,zhang2024ml_bias_e3sm}. Such approaches have been applied to improve simulations of ozone, aerosols, and air-quality fields, as well as to reduce biases in satellite retrievals \citep{ni2026ml_bias_ozone,liu2025deep_learning_ozone,kim2024deep_bcsi_pm25}.
Here, we assess the ability of the ML approach to reproduce EAC4 TCCO trends and annual anomalies. This study represents a first step toward emulating AC data assimilation with machine learning, with the potential of broader applicability to improve the continuity and robustness of long-term atmospheric composition datasets in the presence of observational gaps.

CO is released into the atmosphere by incomplete combustion processes from anthropogenic sources such as road transport and energy production, as well as from wildfires. Secondary production of CO in the atmosphere—primarily from formaldehyde as part of the oxidation chains of $CH_4$, isoprene, and other volatile organic compounds—is of similar or even larger magnitude than direct emissions. Oxidation by OH is the main CO sink, resulting in an atmospheric lifetime of 1–32 months. CO is a key regulator of tropospheric ozone, a short lived climate pollutant.
%This paper investigates whether machine learning can learn the relationship between Copernicus Atmospheric Monitoring Service (CAMS) control run and reanalysis total column carbon monoxide (TCCO) and thereby compensate for the loss of MOPITT observations \citep{Deeter2003MOPITT}. We evaluate a range of machine learning approaches, including linear regression, random forests, gradient boosting, and neural networks. Our results show that a neural network model can reliably reproduce monthly mean reanalysis TCCO using control run TCCO together with additional input features, such as coordinates, month, fire emission and formaldehyde data.

\section{Data}

The dataset used in this study are monthly-mean TCCO fields of a control simulation (without data assimilation) and from CAMS global reanalysis, EAC4 \citep{Inness2019EAC4}. The grid resolution of both data is 0.7 x 0. 7 \textdegree and the original temporal resolution of 6h.  The control simulation and the data assimilation run for EAC4 applied the same meteorological and atmospheric composition model \citep{Flemming2015CIFS} and the same emissions. The main difference between the control run and EAC4 is the application of data assimilation of atmospheric composition satellite retrievals, such as TCCO from the MOPITT instrument.  

\subsection{Input Features}

The input features include total column carbon monoxide (TCCO) from the control run, along with additional variables that may help correct the control run values:

\begin{itemize}
    \item Monthly mean of total column formaldehyde from the control run
    \item Monthly mean of total column isoprene from the control run 
    \item Monthly mean of fire emissions from the Global Fire Assimilation System (GFAS, \citep{Kaiser2012GFAS}), represented by monthly mean, standard deviation, and maximum
    \item Geographical coordinates (latitude and longitude)
    \item Month of the year
\end{itemize}

\subsection{Target}

The single target variable is the total column carbon monoxide from EAC4 reanalysis. This product incorporates data assimilation to combine the control run with satellite observations from MOPITT, providing a more accurate representation of atmospheric CO distribution.

\section{Methods}

To learn the relationship between control run and reanalysis TCCO, we consider four standard, lightweight machine learning (ML) approaches. We choose the best model and hyperparameters using random search, a temporal train validation split, and an ablation study of each input feature. %We note that we do not use latitude-weighting for training or evaluation

\subsection{Data Split for Model Selection}

To select the most promising ML model, in the initial step the data are split temporally into training (2003--2018) and validation (2019--2020) periods. The model with the best validation score will be used as the final model. A temporal split is chosen here, as it provides a realistic assessment of the model's ability to generalize to future periods, which is our target task. All input features are standardized by subtracting the mean and dividing by the standard deviation for each variable. %Each monthly grid point is treated as an independent sample.

\subsection{ML Models}

We consider four commonly used ML methods for tabular data, as we are treating each monthly grid point independently. The models are:  

\begin{itemize}
    \item \textbf{Linear Regression (LR)}
    \item \textbf{Gradient Boosting (GB)} \citep{friedman2001greedy}
    \item \textbf{Random Forest (RF)} \citep{breiman2001random}
    \item A feedforward \textbf{Neural Network (NN)} \citep{rumelhart1986learning}
\end{itemize}

For each method, we define a hyperparameter (HP) grid (see Appendix \ref{app-hps}) and perform a random search with 10 samples to identify the best configuration. In the end, HPs with the lowest validation score are kept. %We note that we 

\begin{table*}[htb] 
\caption{Validation scores for different ML models. Each model is tuned using 10 random samples from the HP grid for each of the different training periods. Best scores are highlighted in bold.} \label{different_models} 
\vskip 0.15in 
\begin{center} 
\begin{small} 
\begin{sc} 
\begin{tabular}{l|cccc|cccc|cccc|cccc} 
\toprule 
Train. Years & \multicolumn{4}{c}{2003-2018} & \multicolumn{4}{c}{2014-2018} & \multicolumn{4}{c}{2017-2018} & \multicolumn{4}{c}{2018}\\ 
\midrule 
Model & LR & GB & RF & NN & LR & GB & RF & NN & LR & GB & RF & NN & LR & GB & RF & NN \\ 
\midrule 
RMSE & .13 & .19 & .09 & .09 & .12 & .11 & .08 & \textbf{.07} & .13 & .18 & .09 & .08 & .12 & .11 & .09 & .10 \\ 
$R^2$ & .93 & .84 & .97 & .96 & .94 & .95 & .97 & \textbf{.98} & .93 & .95 & .97 & .97 & .94 & .95 & .97 & .96\\ 
\bottomrule 
\end{tabular} 
\end{sc} 
\end{small} 
\end{center} 
\vskip -0.1in 
\end{table*}

\subsection{Model and Training Period Selection}

Validation scores for the best model of each method are shown in Table \ref{different_models}. We consider both the $R^2$ score (see Appendix \ref{app-r2} for definition) and root mean squared error (RMSE) \footnote{We note that we do not perform any latitude-weighting for training losses or validation metrics here.}.

\paragraph{ML models}
Overall the neural network gives the best results on the validation scores, closely followed by the random forest approach. This shows that the more complex models here give better skill than a simple approach such as linear regression. Gradient boosting shows mixed results depending on the training period, sometimes worse sometimes better than linear regression.

\paragraph{Training years} In addition to comparing different ML methods and their hyperparameters, we investigate the impact of the training period length. We test models trained on all available years up to the year before validation, as well as on 5-year, 2-year, and 1-year periods. The results indicate that the NN trained on 2014--2018 is the most promising approach with the lowest RMSE of $0.07$ molec/cm$^2$ and a $R^2$ score of $0.98$. The 5-year training period performing best might be due to a trade-off between amount of training data used and the recency of that data.

\subsection{Ablation of Input Features}

\begin{table*}[htb] 
\caption{Ablation results: -- The hyperparameter-tuned NN is trained again dropping one input features at a time. Best scores are highlighted in bold}
\label{ablate_nn}
\vskip 0.15in
\begin{center}
\begin{small}
\begin{sc}
\begin{tabular}{l|cccc}
\toprule
Ablate & RMSE & $R^2$ \\
\midrule
none & \textbf{.074} &  \textbf{.977} \\
\midrule
TCCO CR & .158 & .895 \\
Coordinates & .101 & .957  \\
Month & .098 &  .960 \\
Isoprene (CR) & \textbf{.074} &  \textbf{.977}\\
Formaldehyde (CR) & .075 & .976\\
GFAS mean & .075 & \textbf{.977}  \\
GFAS std &  .075 &  \textbf{.977} \\
GFAS max &  .075 & \textbf{.977} \\
\end{tabular}
\end{sc}
\end{small}
\end{center}
\vskip -0.1in
\end{table*}

We perform an ablation study to assess the contribution of each input feature (see Table \ref{ablate_nn}). Here, we find that including isoprene does not improve performance -- measured in RMSE -- so it is removed from the final model. The permance drop for each variable is also an indicator for the importance of that feature, we can observe that apart from the CR TCCO itself coordinates and month have the strongest influence on the ML prediction, whereas fire emissions and formaldehyde, as proxy for the chemical production of CO,  have a small contribution. 

\subsection{Different Normalization Techniques}

To see if we can further improve performance, especially with respect to the application case of anomaly maps, we apply an additional normalization on top of the the initial z-scaling of input features. Both input and target CO are standardized with the mean and standard deviation of their respective location, calculated on the 5 year training period of the EAC4 target CO -- a kind of map normalization. We also keep the initial version and call these two approaches ML and $\text{ML}_{\text{ano}}$. $\text{ML}_{\text{ano}}$ reaches an RMSE of $0.12$ and a $R^2$ score of $0.94$ on the validation years 2019-2020. This is significantly worse than the initial option, but we will still keep it as it can improve anomaly map prediction skill.

\subsection{Production of Results}

For a detailed evaluation of the ML performance we produce results for the whole period ranging from 2004 to 2025. For each target year, the ML correction is produced by training an NN (HPs are specified in Appendix \ref{app-hps}) on the preceding five years. That way each NN for each year has the same HPs but different model weights. If fewer than five years of data are available -- which is the case for 2004 -- 2007 -- all available previous years are used. Each resulting model is then evaluated on the year immediately following its training period, as described in the next section.

\begin{figure*}[htb]
\includegraphics[width=14cm]{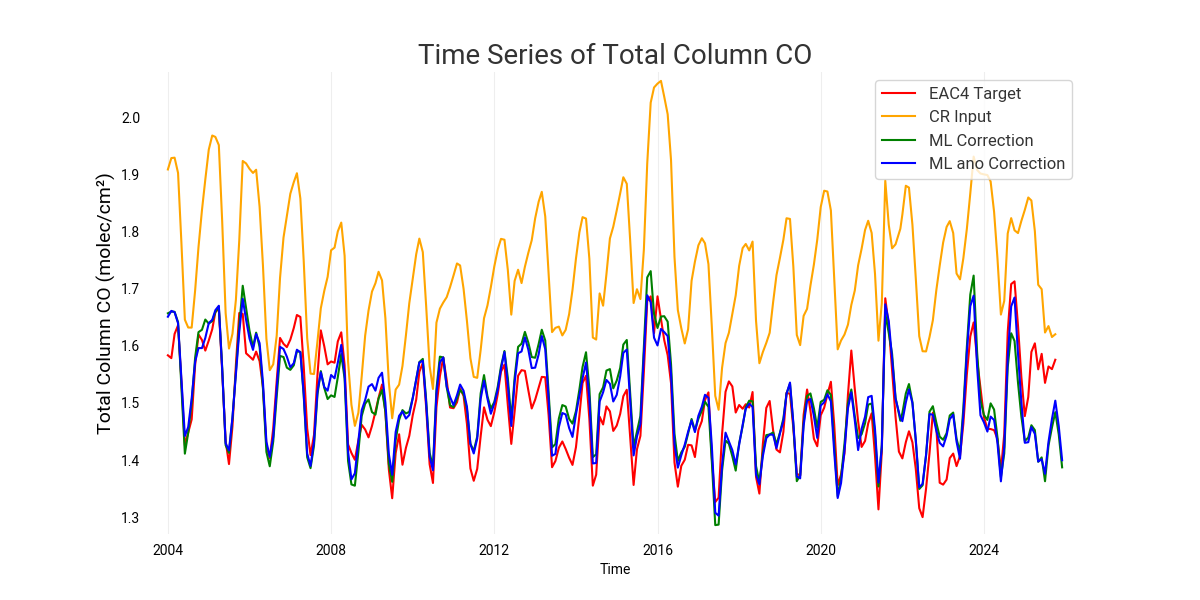} 
\caption{The global means timeseries of CR CO (Input), EAC4 CO (Target), ML and $\text{ML}_{\text{ano}}$ CO (prediction) for the whole dataset period. For each year the ML model is trained on the previous (max.) 5 years.}
\label{timeseries}
\end{figure*}

\section{Results}

Figure \ref{timeseries} presents the global mean TCCO time series from 2004 to 2025, where the ML model is retrained each year using the preceding five years of data. The ML-corrected TCCO closely follows the EAC4 reanalysis throughout the entire period and shows a substantial improvement compared to the control run (CR). The CR exhibits a pronounced positive bias relative to the reanalysis, which is effectively reduced by the ML corrections in all years. In 2025, EAC4 diverges from the ML predictions because no MOPITT TCO observations are assimilated in this period. The ML prediction in 2025 is more consistent with the EAC4 TCCO field for the period 2004-2024.  

\begin{figure*}[htb]
\includegraphics[width=15cm]{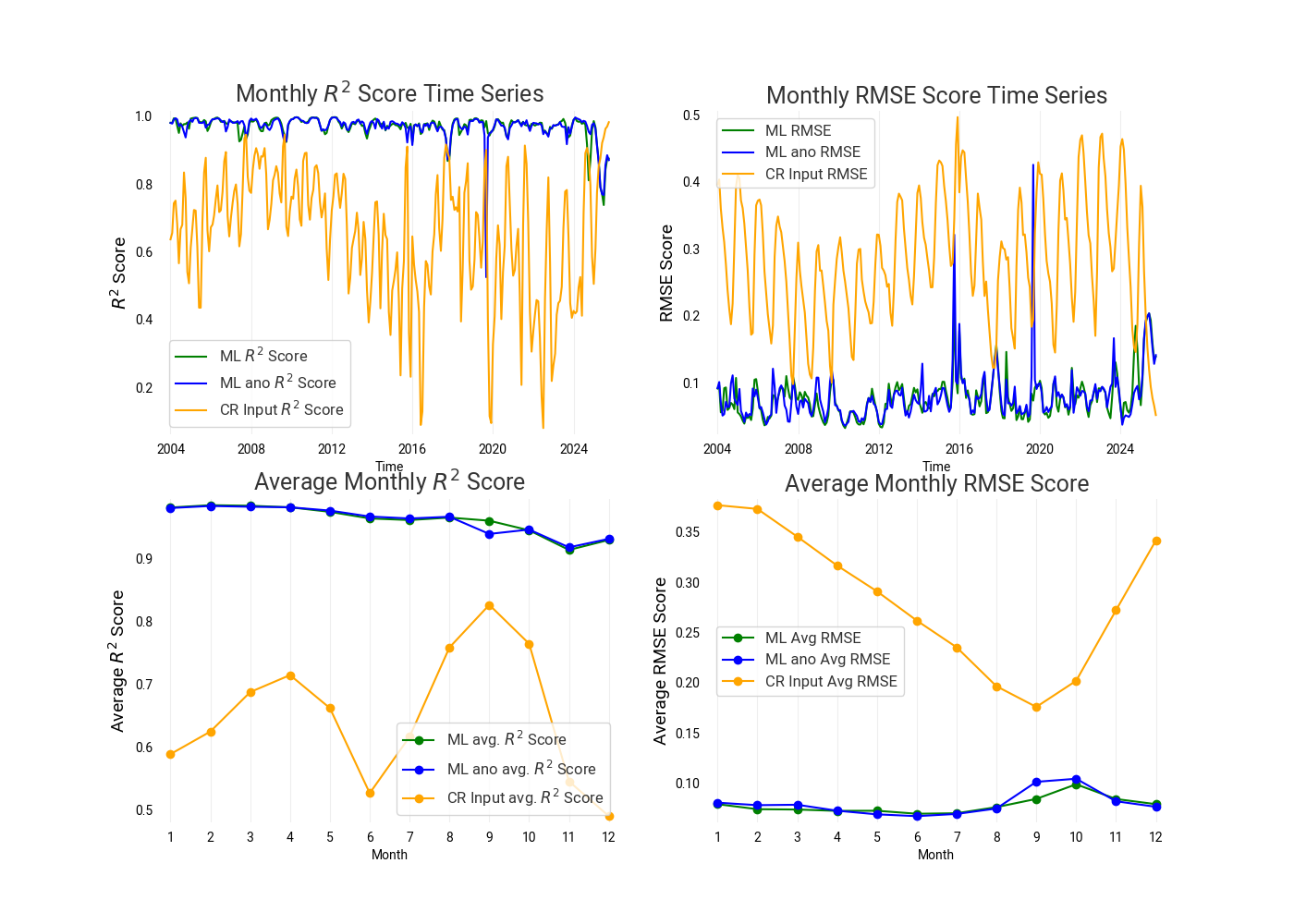}
\caption{Top row: Monthly $R^2$ scores and RMSE between the different datasets over the full period. Bottom row: Mean monthly $R^2$ scores and RMSE, highlighting seasonal variations.}
\label{score_timeseries}
\end{figure*}

Model performance is further quantified using RMSE and the coefficient of determination ($R^2$). The second row of Figure \ref{score_timeseries} shows the averaged per month of these metrics. The differences between the CR and the reanalysis exhibit a clear seasonal cycle. The RMSE reaches a minimum in September and peaks in January and February. The $R^2$ score displays two annual maxima, with the highest agreement in September and a secondary peak in May, while the largest discrepancies occur in December.

In contrast, the ML-corrected TCCOs show substantially reduced seasonal variability in both RMSE and $R^2$, indicating that the model successfully compensates for systematic seasonal differences between the CR and the reanalysis. Only a weak increase in RMSE is observed in October.

\begin{figure*}[htb]
\includegraphics[width=12cm]{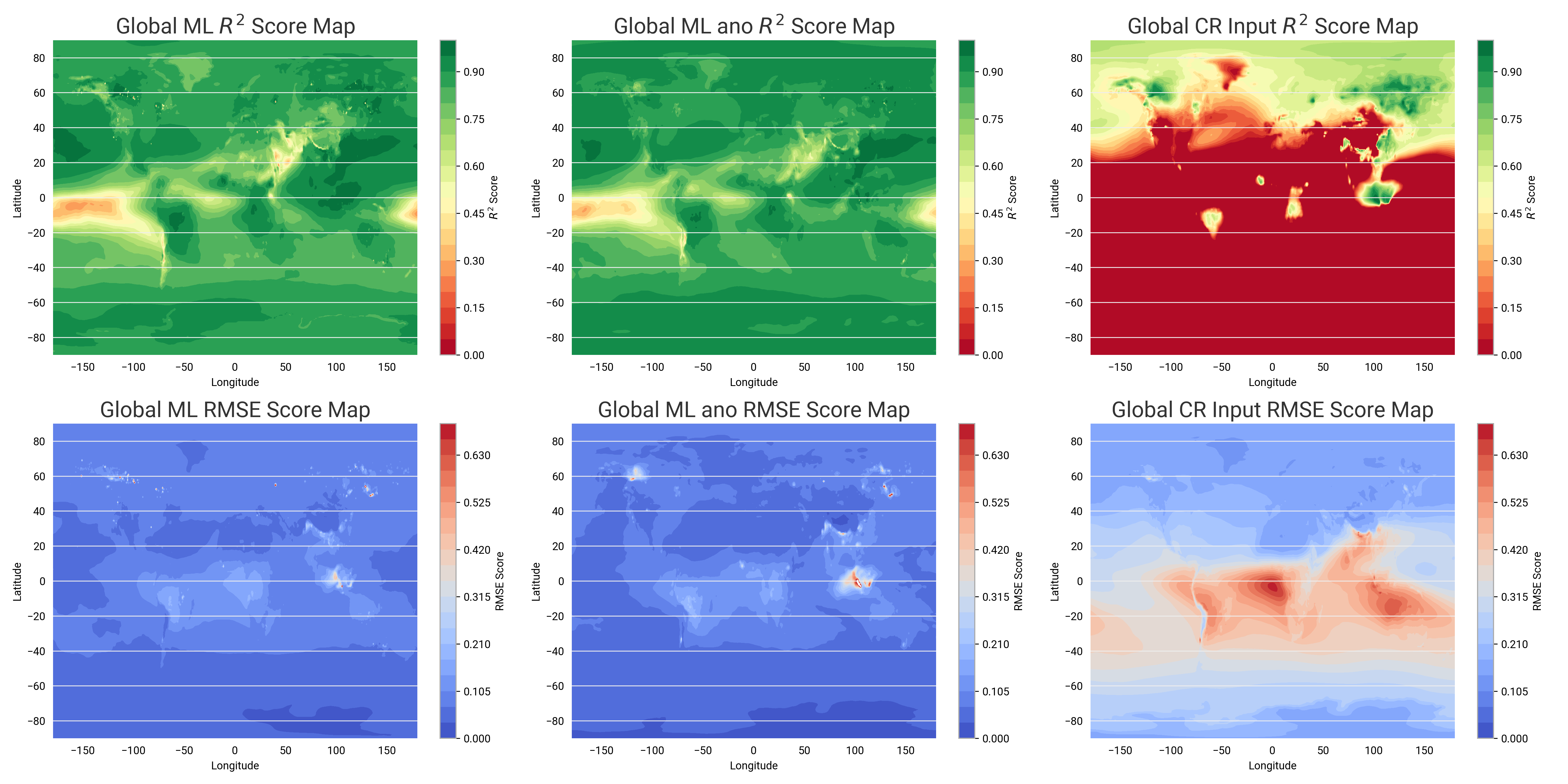}
\caption{This figure shows the scores for each grid point over the yeas from 2004-2024. The left column shows the ML scores and the right column shows the CR scores.}
\label{score_map}
\end{figure*}

The spatial distributions of $R^2$ and RMSE are shown in Figure \ref{score_map}. The ML corrections improve agreement with the reanalysis across all regions globally. In particular, large discrepancies present in the CR, especially in tropical regions, are strongly reduced by the ML model, demonstrating its ability to capture spatially heterogeneous biases.

To investigate both spatial and seasonal variability we show monthly differences of the ML corrections and CR compared to EAC4 in the appendix, see Figure \ref{monthly_maps}, for the year 2024. This shows again how the strong positive bias of CR, apparent globally in all seasons, is corrected with both ML models. In the months of September until December we can observe that the standard ML approach exhibits a slight negative bias in the southern hemnisphere, which is less apparent when using the map normalization under $\text{ML}_\text{ano}$.

\subsection{Application: Anomaly Maps}

A key application of atmospheric composition reanalyses is the generation of yearly anomaly maps, which are widely used to diagnose interannual variability and extreme events. Here, we assess the performance of the two ML model variants in reproducing TCCO anomaly patterns (see Appendix for corresponding figures).

To quantify the quality of the predicted anomaly maps, we use a sign alignment metric, defined as the fraction of grid points for which the predicted and reference anomalies have the same sign, thereby measuring the consistency of positive and negative deviations from the mean. Using this metric, the normalized anomaly-based ML approach ($\text{ML}_{\text{ano}}$) outperforms the standard ML correction, achieving a score of $0.84$, compared to $0.77$ for the standard ML model and $0.74$ for the control run. This is an indication that the ML correction not only removes constant biases but corrects more complex patters. 

In addition, the average RMSE of the anomaly fields is reduced from $0.066$ for the control run to $0.043$ for the standard ML correction, and further to $0.041$ for $\text{ML}_{\text{ano}}$. These results indicate that incorporating anomaly normalization improves the model’s ability to capture both the magnitude and spatial structure of interannual variability.

\section{Conclusions}

In this study, we demonstrate that neural networks can successfully learn the relationship between a CAMS control simulation and the EAC4 atmospheric composition reanalysis and can be used to compensate for changes or losses in assimilated observations. Trained on historical periods, the ML approach accurately reproduces reanalysis total column carbon monoxide (TCCO) for the year following the training period, substantially reducing the systematic positive bias present in the control run. Improvements are observed in both global mean time series and spatial patterns, as well as in the representation of annual anomalies.

The ML corrections effectively captures seasonal variations and regional discrepancies between the control run and the reanalysis, leading to consistently higher $R^2$ scores and lower RMSE values across time and space. In particular, the anomaly-based normalization further enhances the skill of the ML model in reproducing interannual variability, making it well suited for applications such as anomaly monitoring in periods affected by observational discontinuities.

These results highlight the potential of machine learning as a complementary tool to traditional data assimilation for maintaining temporal consistency in long-term atmospheric composition datasets. While the present study focuses on monthly mean TCCO, the approach is general and could be extended to other trace gases and aerosol-related variables within the CAMS framework.

Future work could explore the use of spatially aware ML architectures, such as convolutional neural networks or vision transformers, to explicitly model spatial dependencies and potentially improve performance at finer temporal resolutions. Extending the methodology beyond monthly means to daily or sub-daily timescales, as well as assessing predictive skill further into the future beyond one-year lead times, would also be valuable directions for further research.  

Overall, this study provides a first step toward leveraging machine learning to emulate the impact of observational data assimilation, offering a promising pathway for improving the robustness of atmospheric composition reanalyses in the face of evolving satellite observing systems. A fruitful application of the method would be the backward extension of EAC4 into periods without satellite atmospheric composition satellite observations.

%\codeavailability{TEXT} %% use this section when having only software code available

%\dataavailability{TEXT} %% use this section when having only data sets available

%\codedataavailability{TEXT} %% use this section when having data sets and software code available

%\sampleavailability{TEXT} %% use this section when having geoscientific samples available

\newpage

\appendix
\section{}    %% Appendix A

\subsection{Coefficient of determination} \label{app-r2}

The coefficient of determination also called $R^2$ score is defined as follows
\begin{equation}
R^2 = 1 - \frac{\sum_{i=1}^{N} \left( y_i - \hat{y}_i \right)^2}{\sum_{i=1}^{N} \left( y_i - \bar{y} \right)^2}
\end{equation}
where $y_i$ are the reference (reanalysis) values, $\hat{y}_i$ are the model predictions, and $\bar{y}$ is the mean of the reference values. An $R^2$ score of 1 would represent a perfect prediction, 0 represents no skill. Note that the values can be negative too, here they are often set to 0 for better interpretability.

\subsection{Hyperparameters} \label{app-hps}    %% Appendix A1, A2, etc.
The different hyperparameter grids are the following for random forest, gradient boosting and neural networks. The machine learning models are provided by scikit-learn and if not indicated otherwise, the parameters are kept as scikit-learn defaults.

For the random forest the grid consists of
\begin{itemize}
    \item Number of trees: [20,50,100]
    \item Maximum depth: [10,50,100]
    \item Maximum samples: [0.3, 0.5, 0.8].
\end{itemize}

For gradient boosting the grid consists of
\begin{itemize}
    \item Number of trees: [50,100, 200]
    \item Maximum depth: [10,50,100]
    \item Maximum samples: [0.3, 0.5, 0.8].
    \item Learning rate: [0.01, 0.005, 0.001].
\end{itemize}

For the neural network the grid consists of
\begin{itemize}
    \item Hidden layer size: [(100), (100, 50), (150, 75)]
    \item Learning rate: [0.01, 0.001, 0.0005].
\end{itemize}

The final used hyperparameters for the neural network are: a hidden layer size of (100, 50) and a learning rate of 0.001.

%\subsection{Ablation results} \label{app-abl}

%\subsection{Monthly Maps}

%In Figure \ref{monthly_maps} we show the difference maps between the ML versions applied to 2024 and EAC4 as well as the difference between CR and EAC4. This is done for each month of the year.

\begin{figure*}[htb]
\includegraphics[width=14cm]{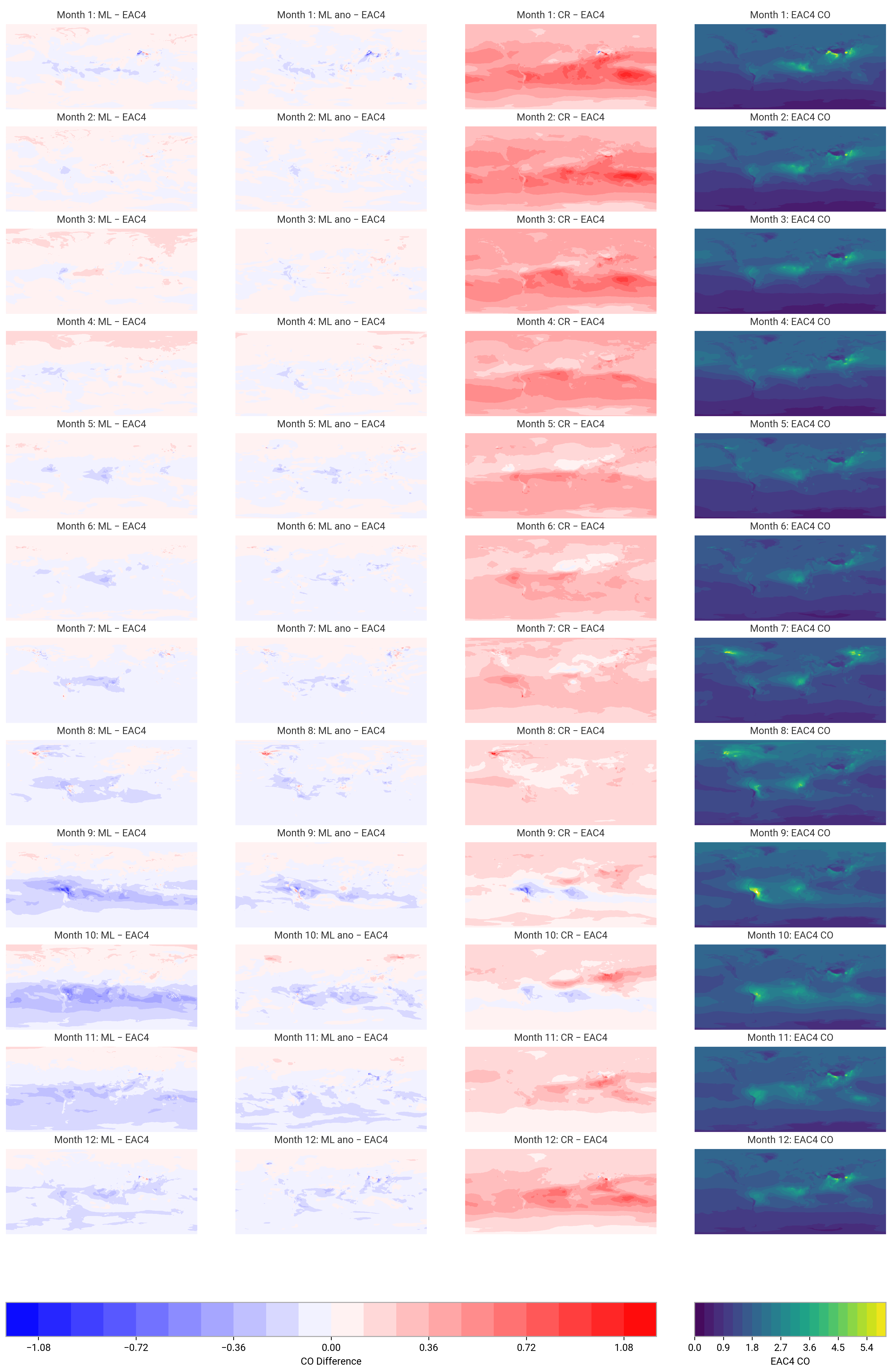}
\caption{This figure shows the spatial error for ML prediction and CR for 2024. The NN is trained on years 2019--2023}
\label{monthly_maps}
\end{figure*}

%\subsection{Anomaly Map Plots}

\begin{figure*}[htb]
\includegraphics[width=14cm]{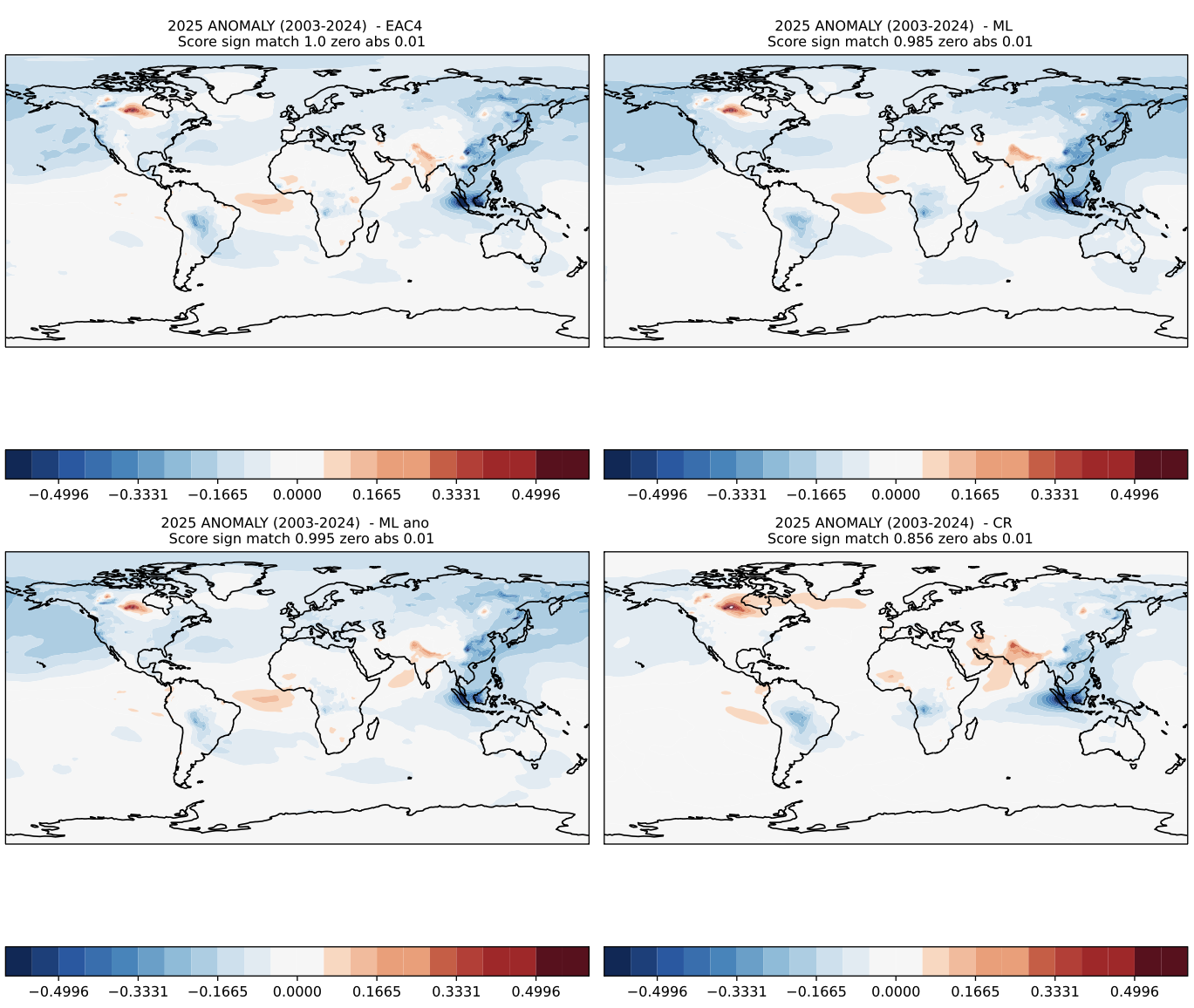}
\includegraphics[width=14cm]{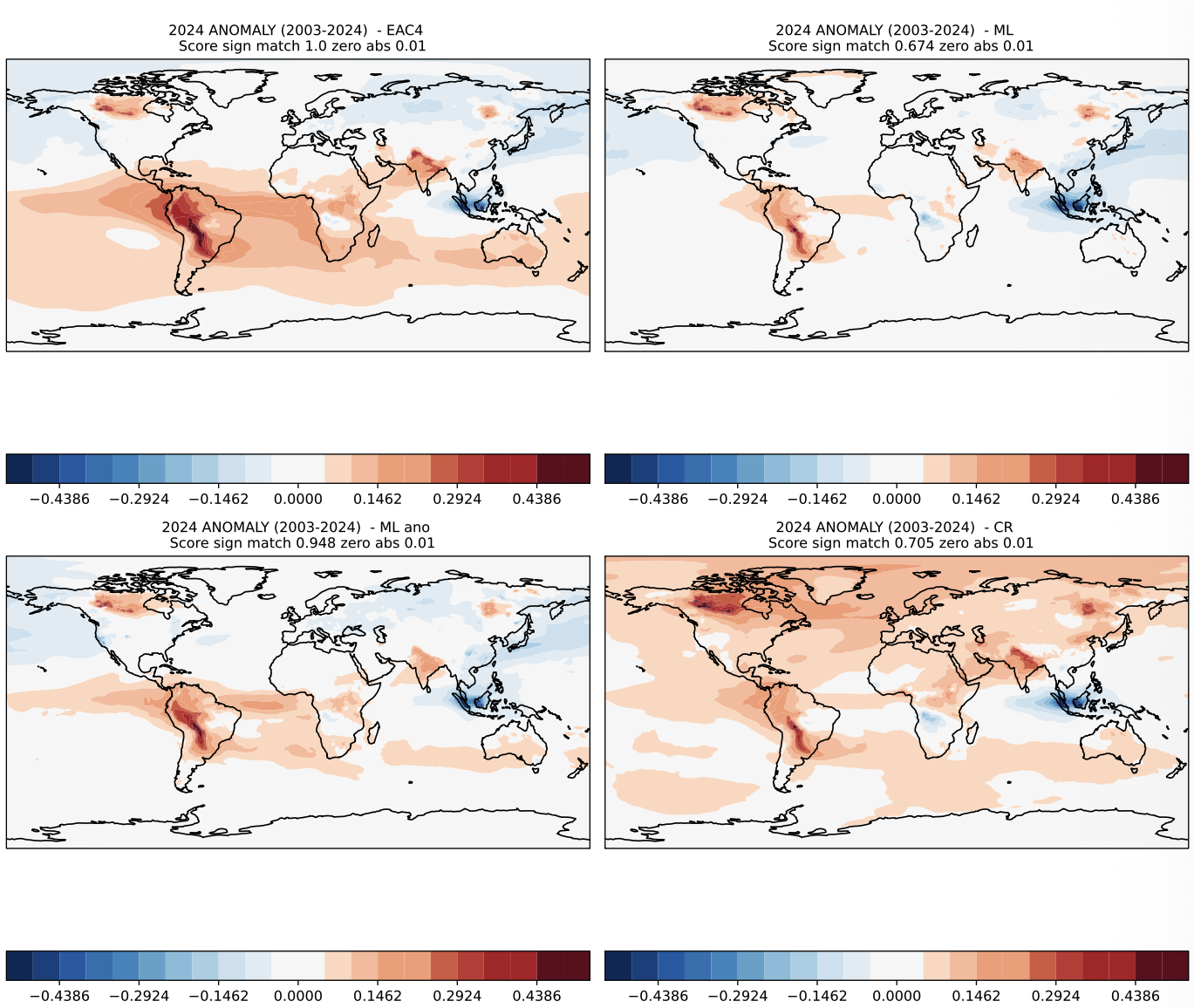}

\caption{.}
\label{monthly_maps}
\end{figure*}

\begin{figure*}[htb]
\includegraphics[width=14cm]{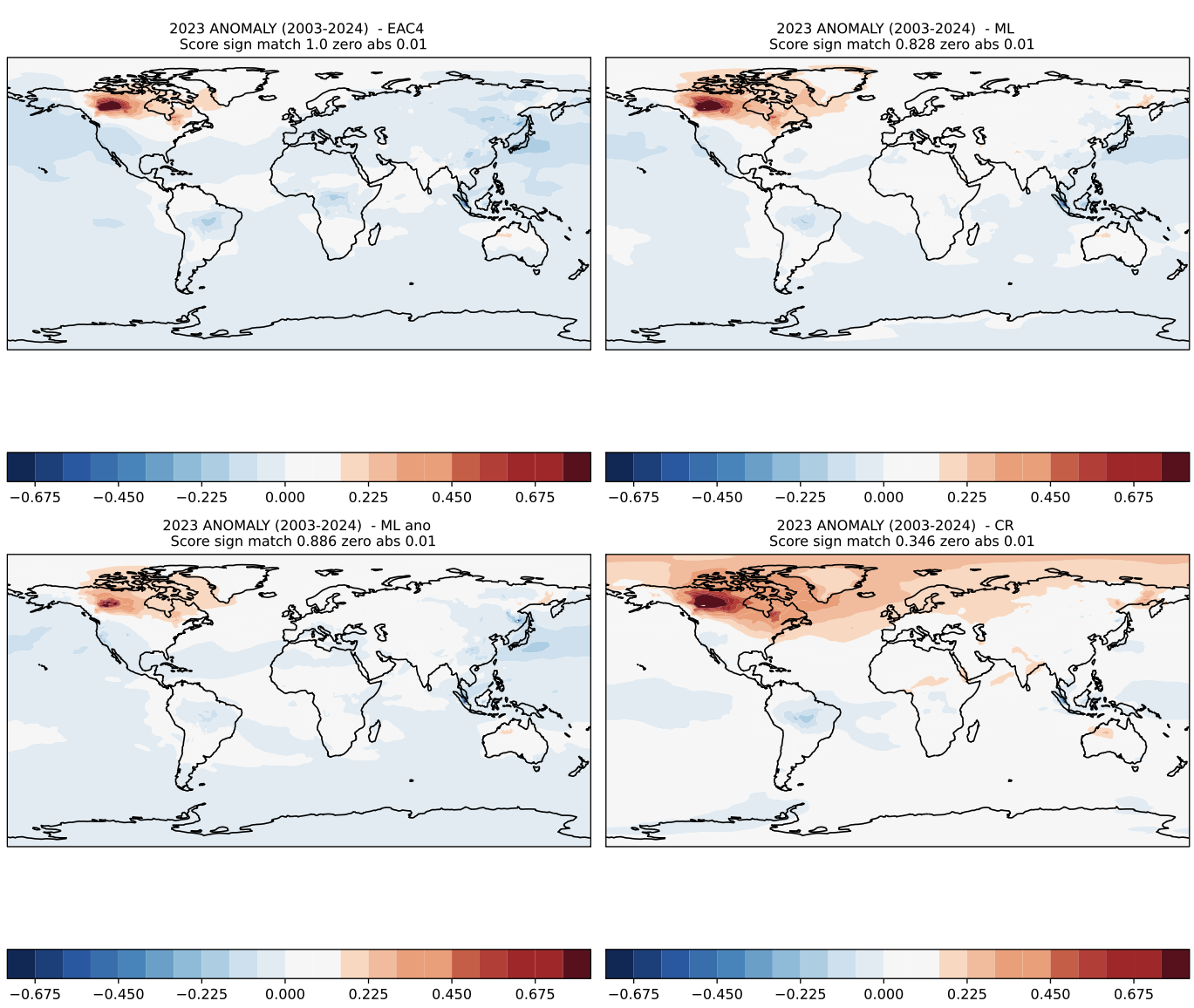}
\includegraphics[width=14cm]{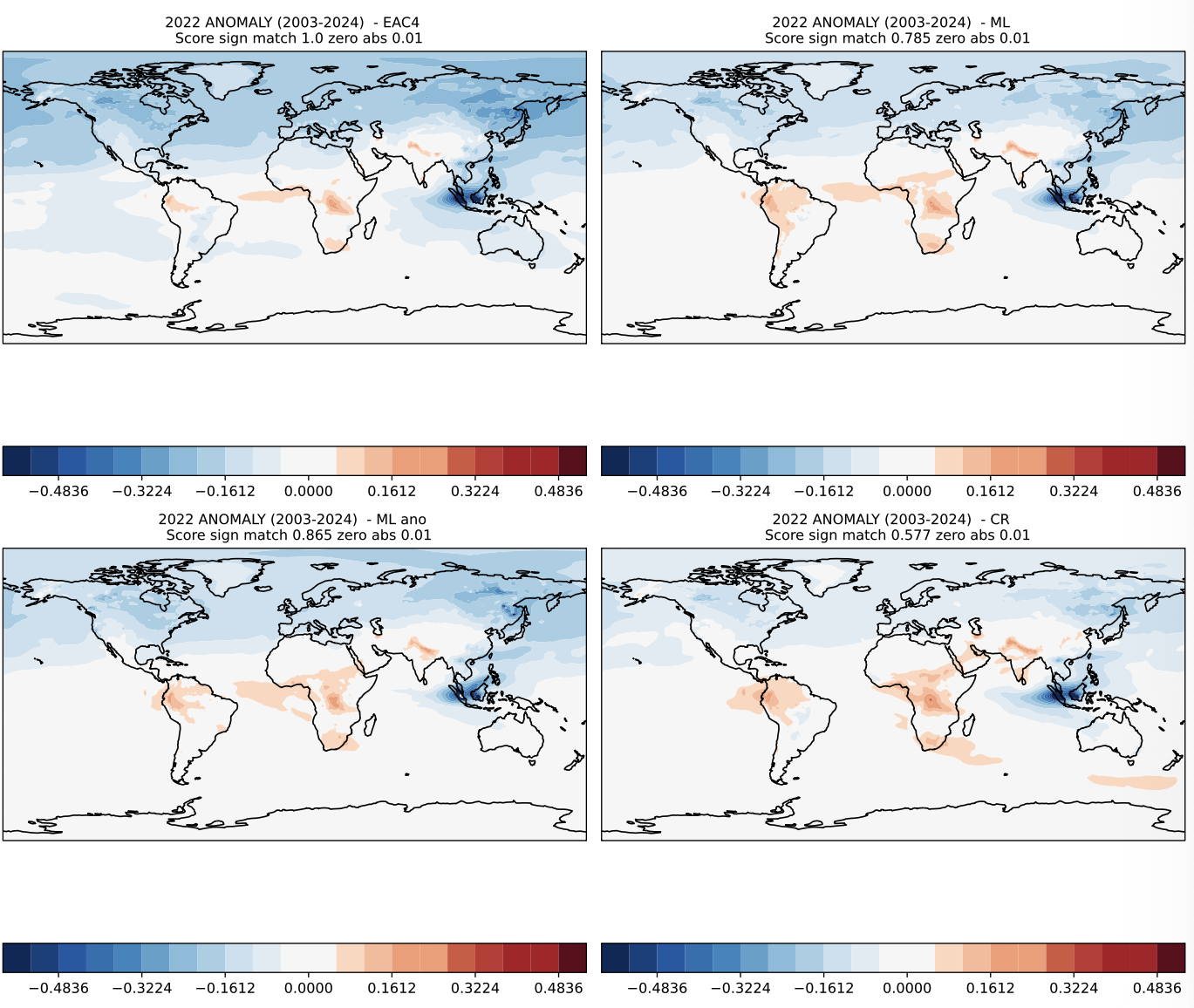}

\caption{.}
\label{monthly_maps}
\end{figure*}

\begin{figure*}[htb]
\includegraphics[width=14cm]{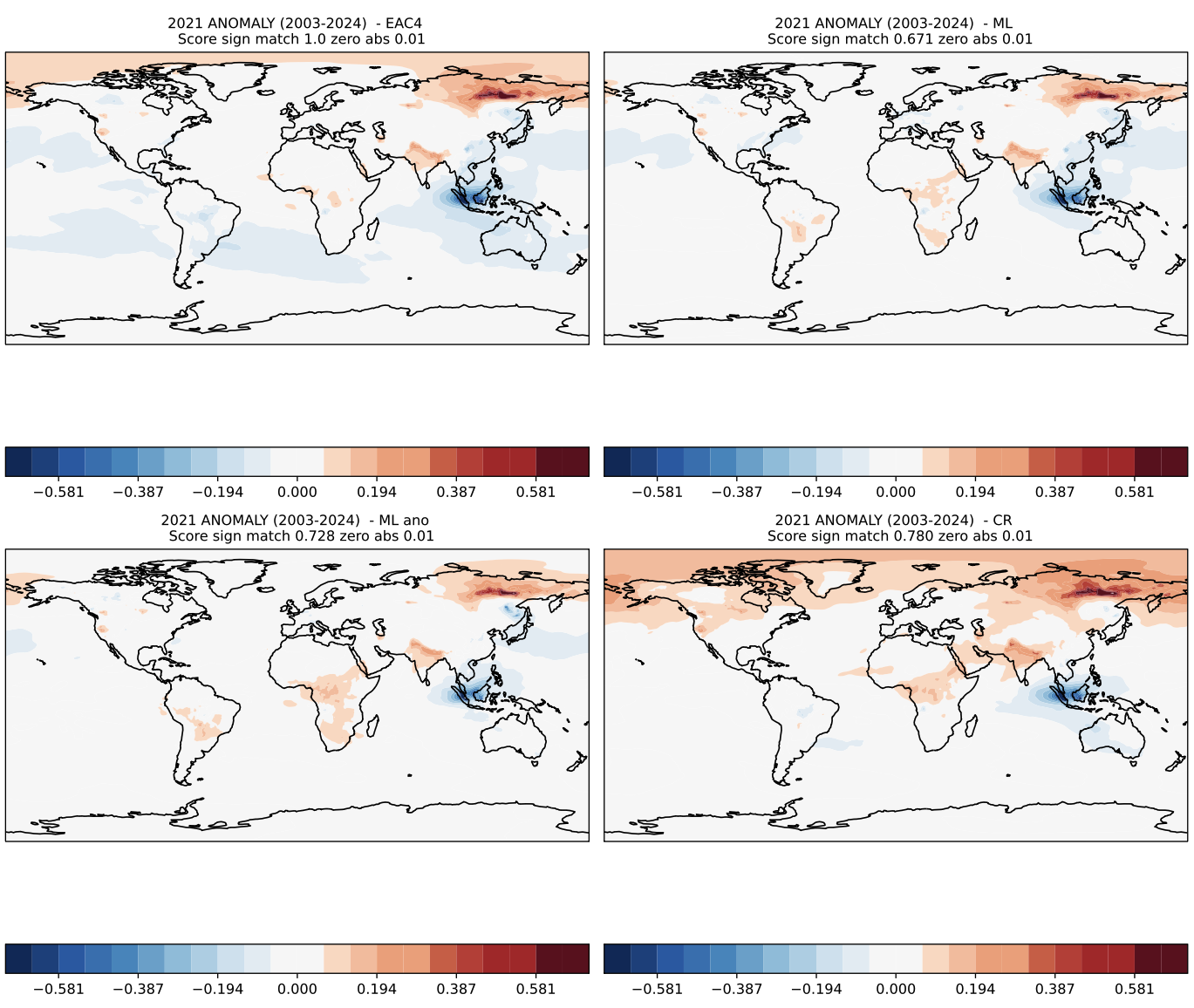}
\includegraphics[width=14cm]{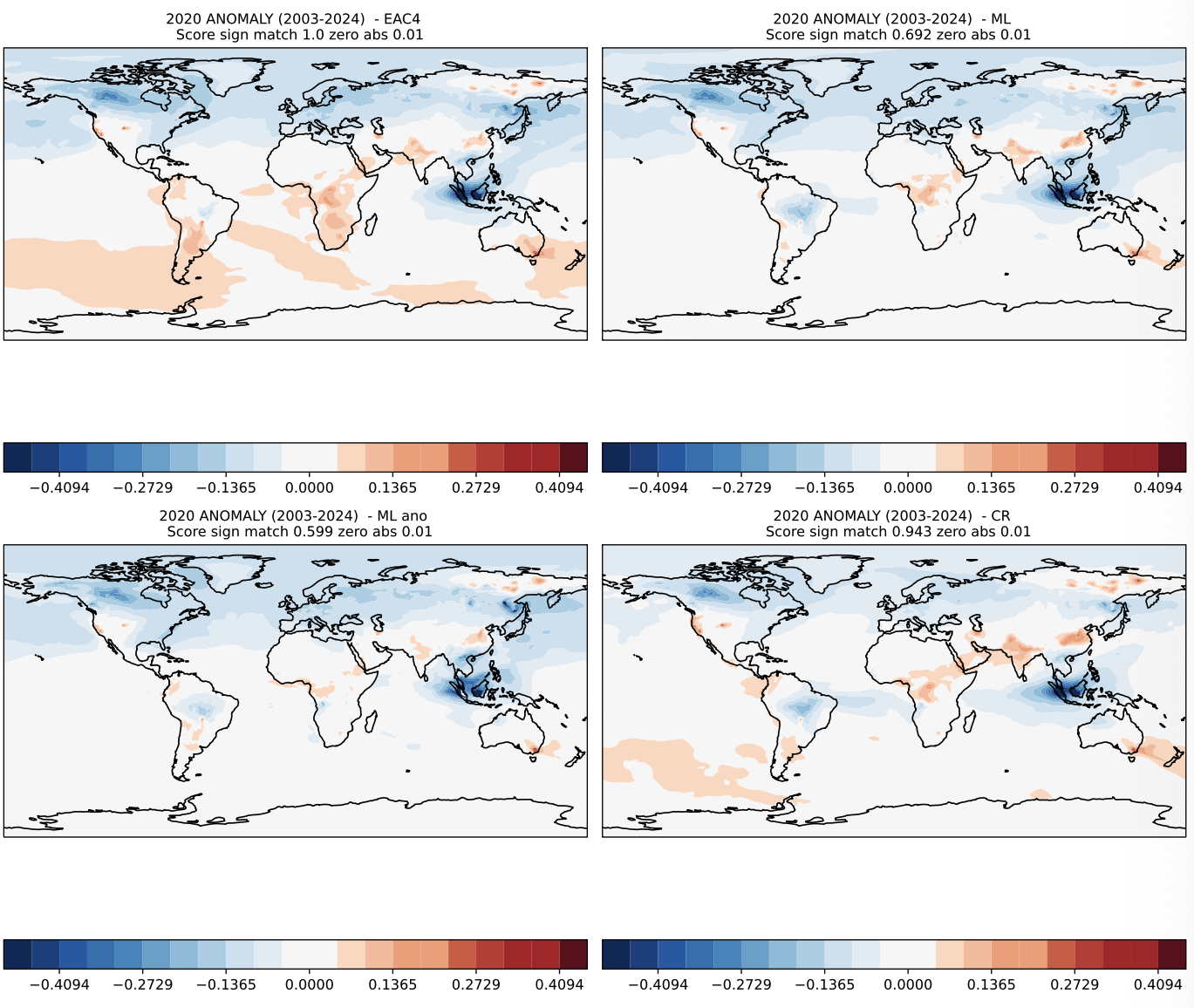}

\caption{.}
\label{monthly_maps}
\end{figure*}

\begin{figure*}[htb]
\includegraphics[width=14cm]{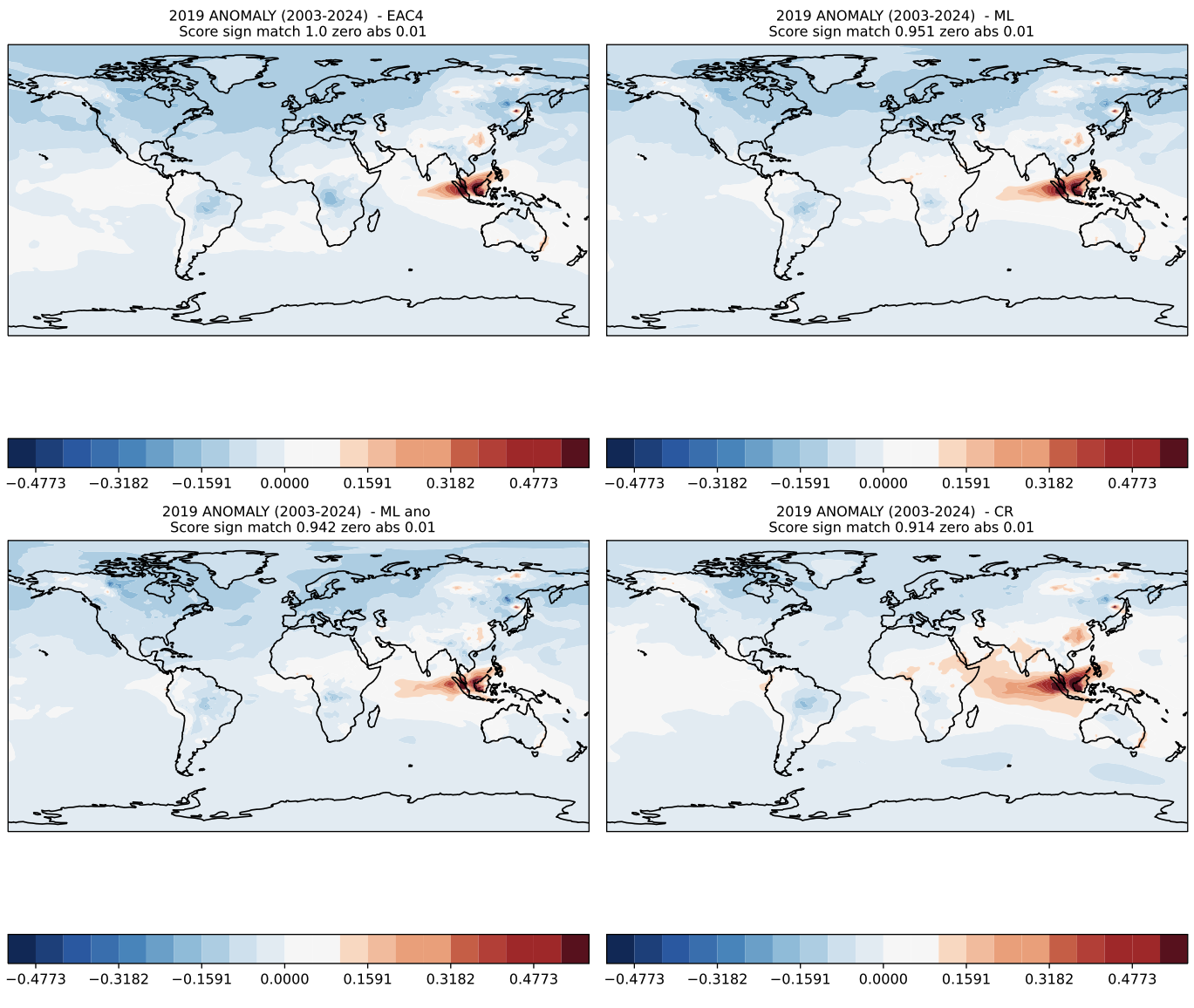}
\includegraphics[width=14cm]{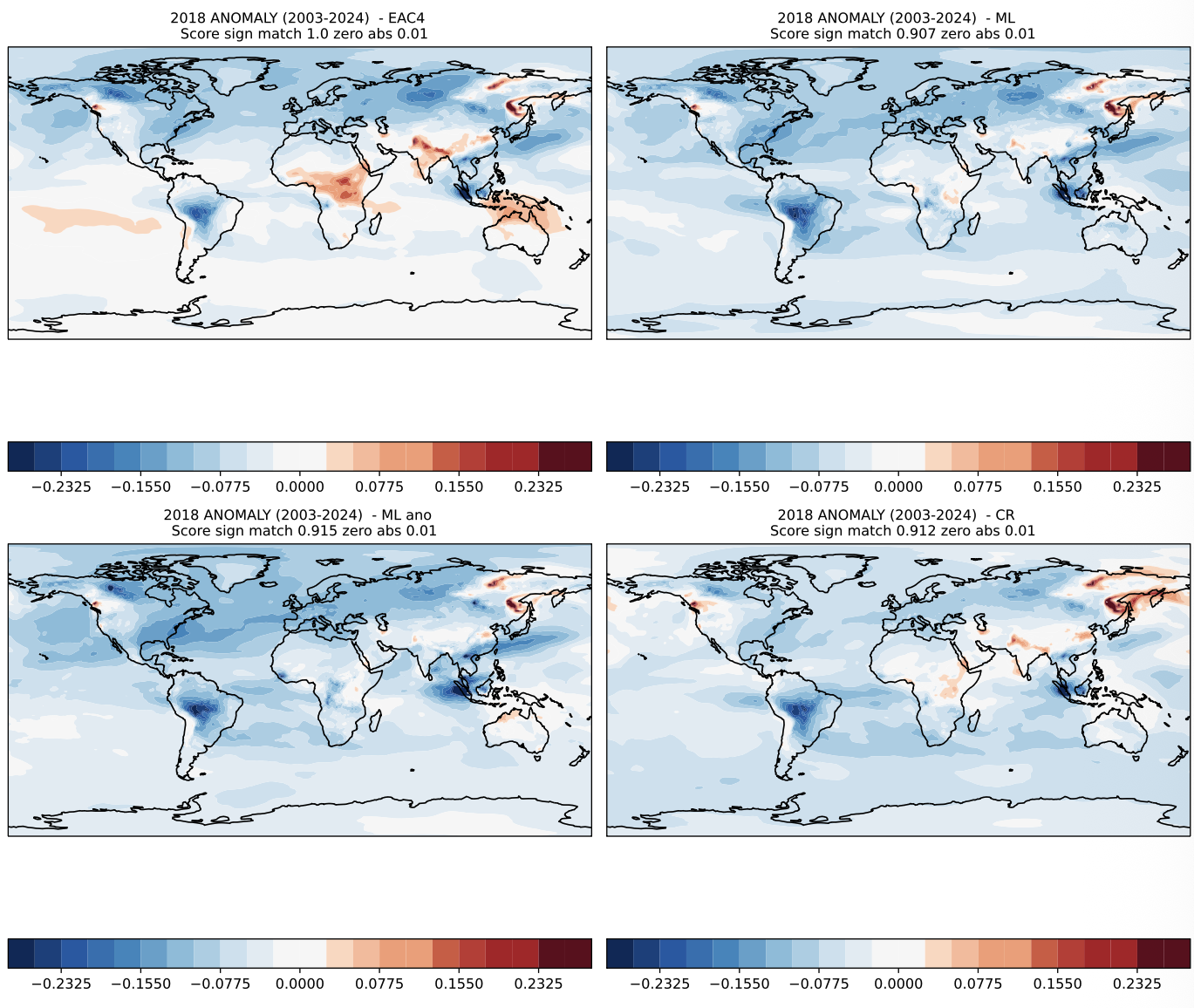}

\caption{.}
\label{monthly_maps}
\end{figure*}

\begin{figure*}[htb]
\includegraphics[width=14cm]{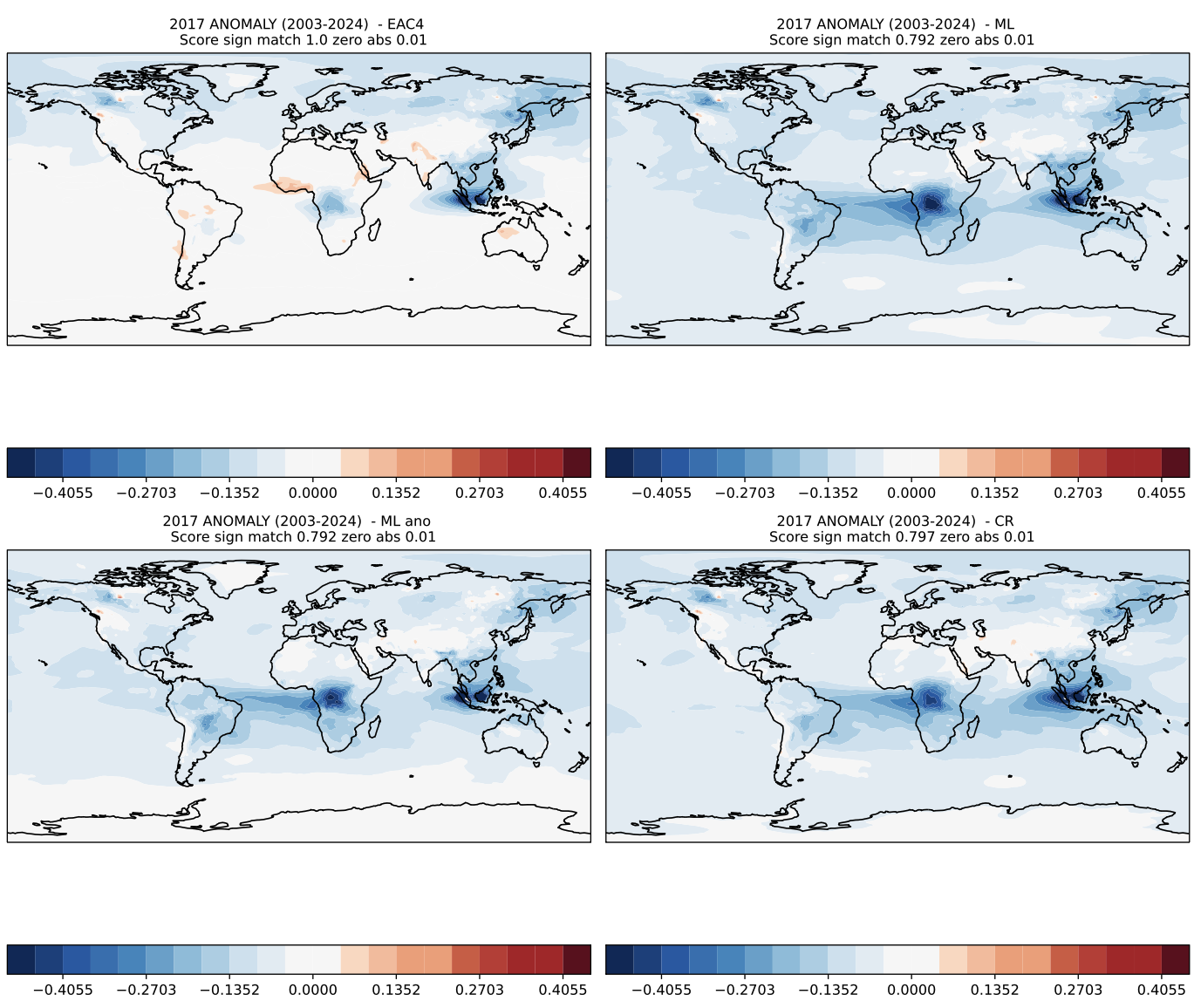}
\includegraphics[width=14cm]{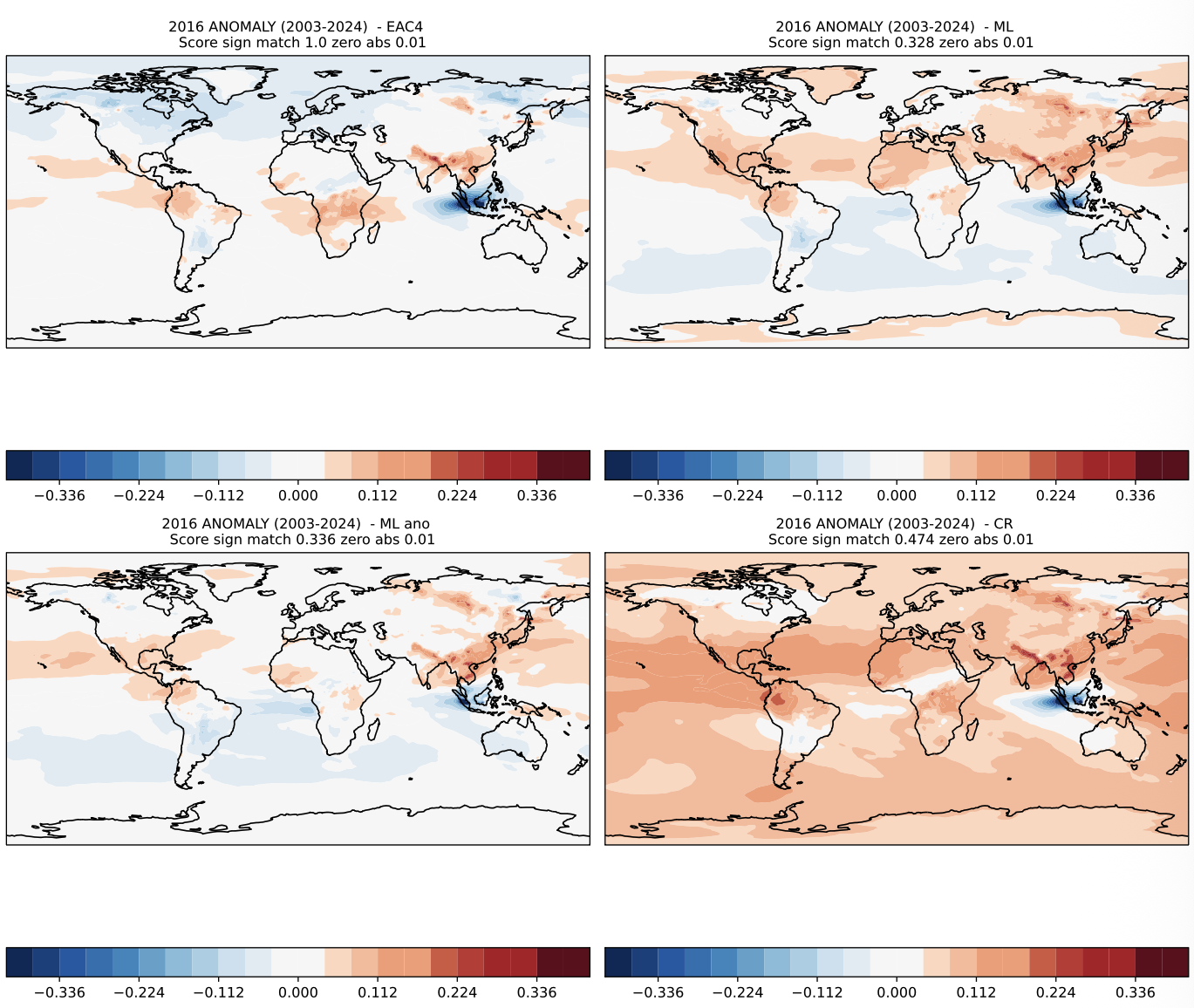}

\caption{.}
\label{monthly_maps}
\end{figure*}

\begin{figure*}[htb]
\includegraphics[width=14cm]{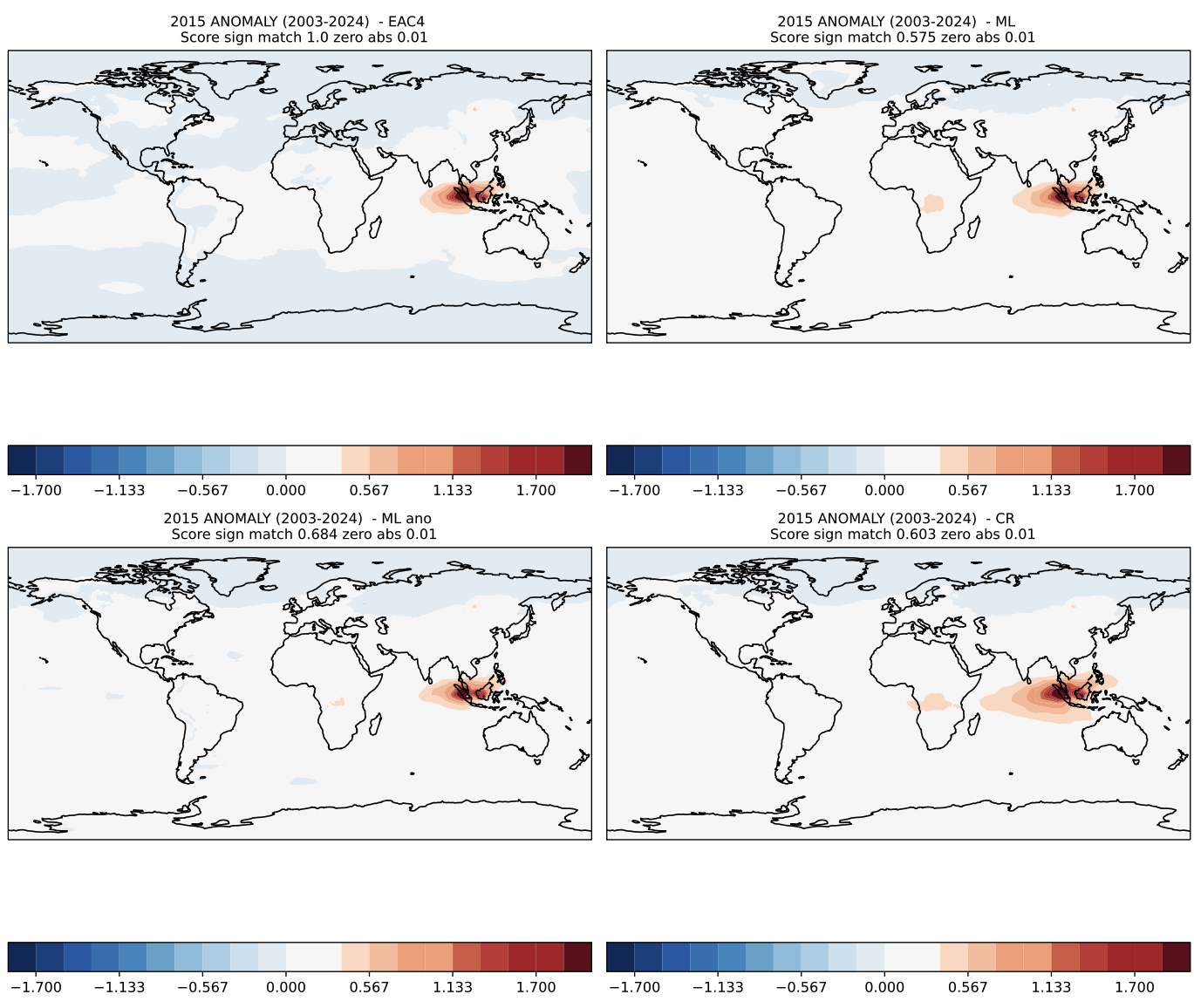}
\includegraphics[width=14cm]{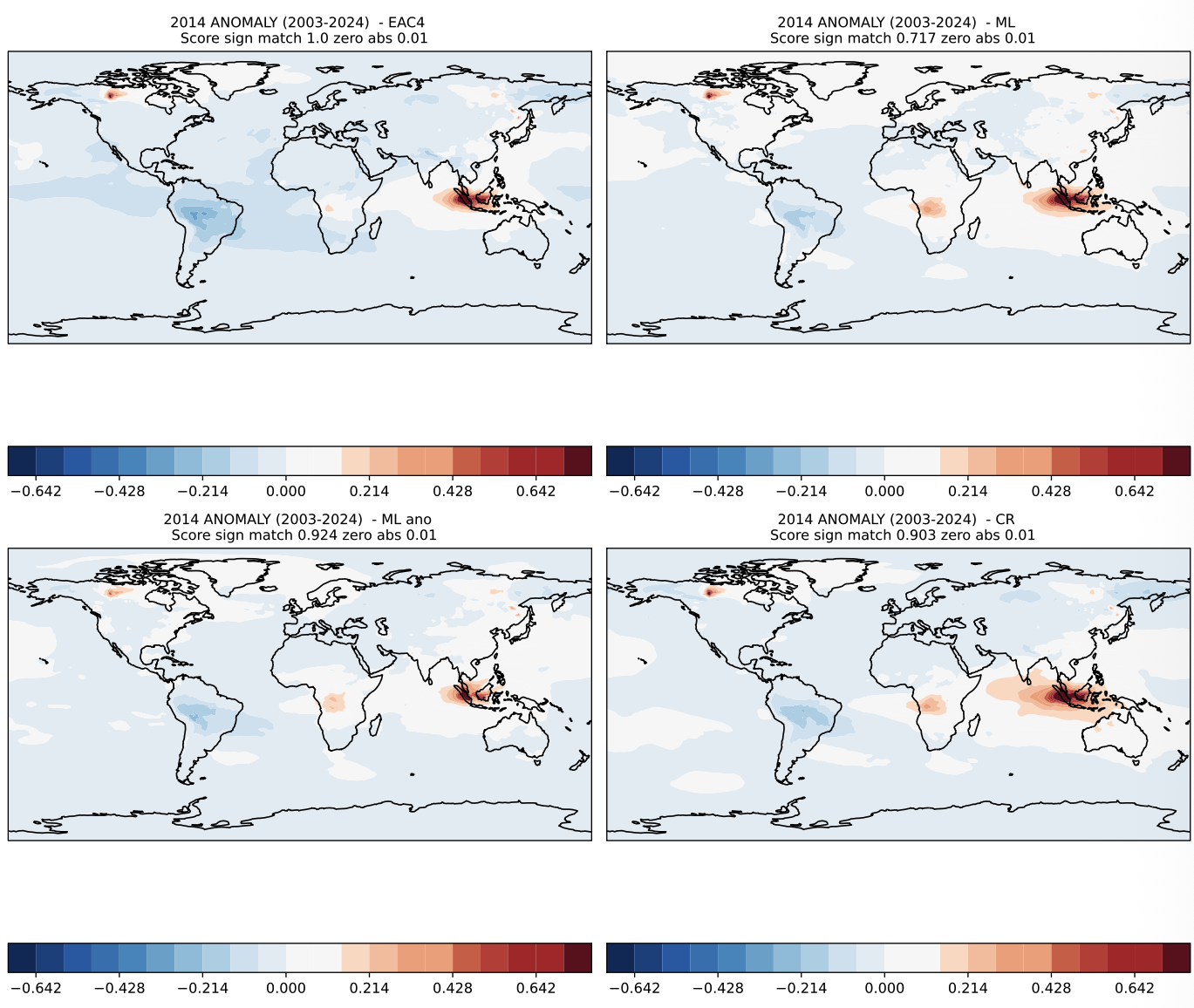}

\caption{.}
\label{monthly_maps}
\end{figure*}

\begin{figure*}[htb]
\includegraphics[width=14cm]{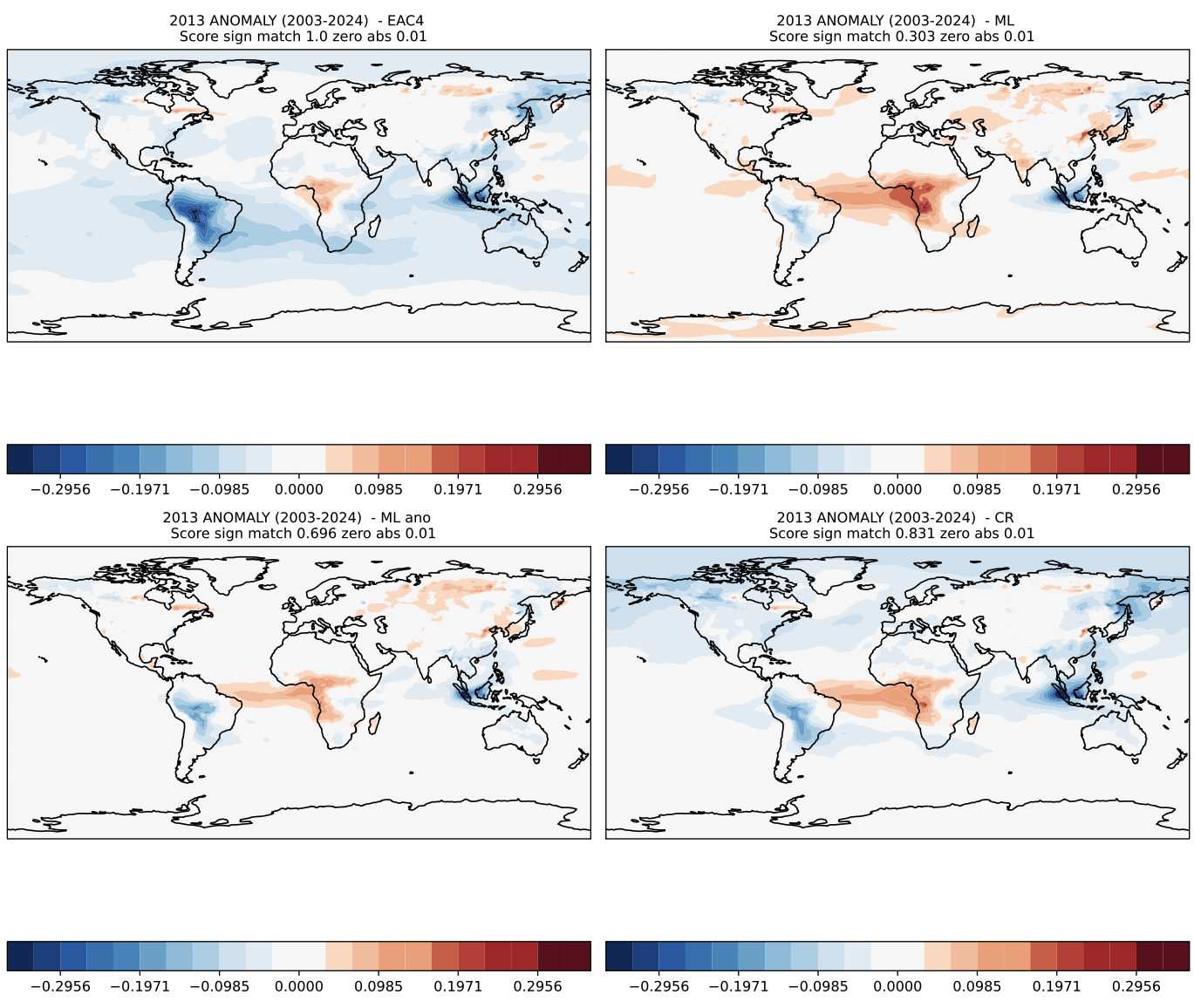}
\includegraphics[width=14cm]{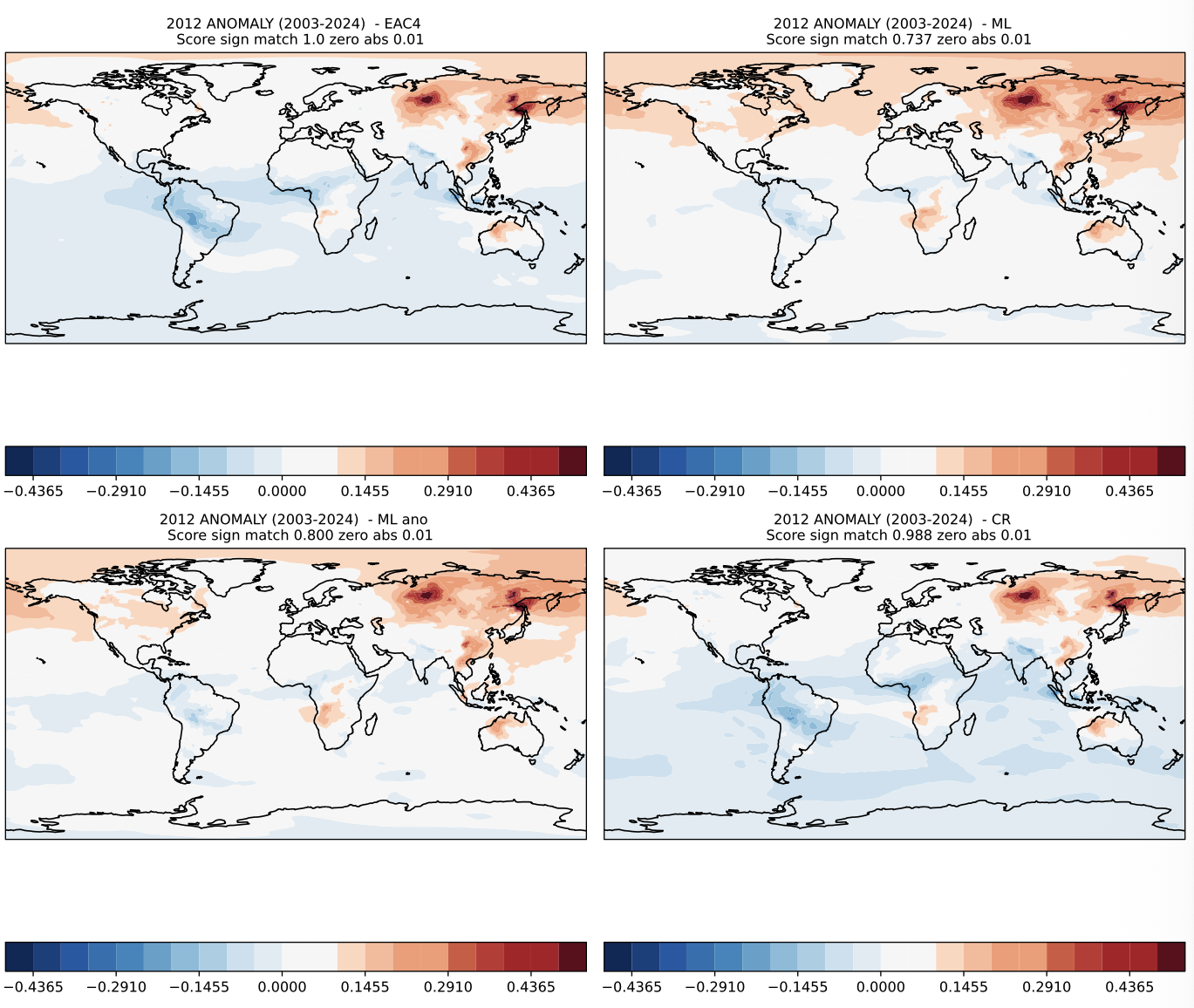}

\caption{.}
\label{monthly_maps}
\end{figure*}

\begin{figure*}[htb]
\includegraphics[width=14cm]{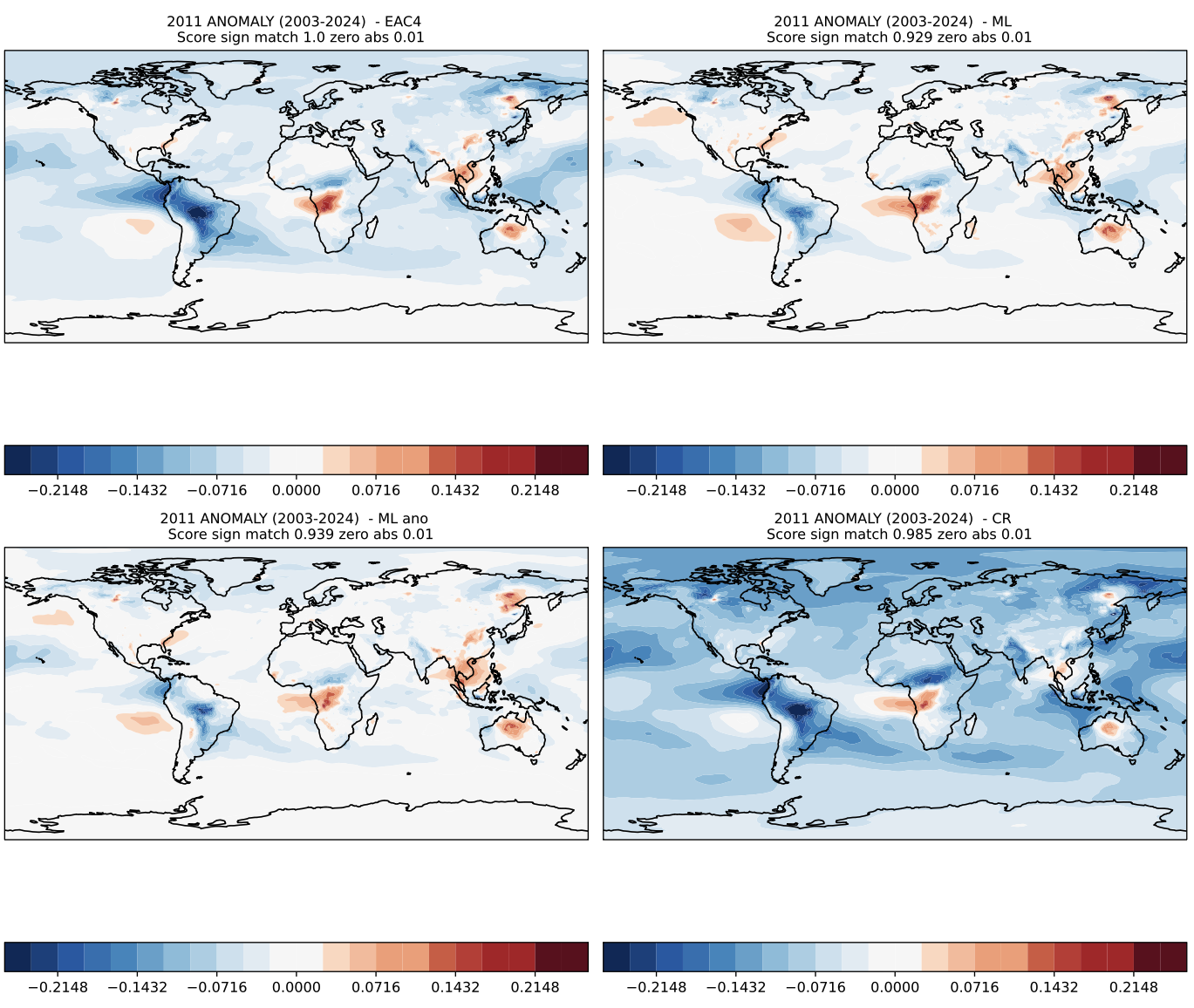}
\includegraphics[width=14cm]{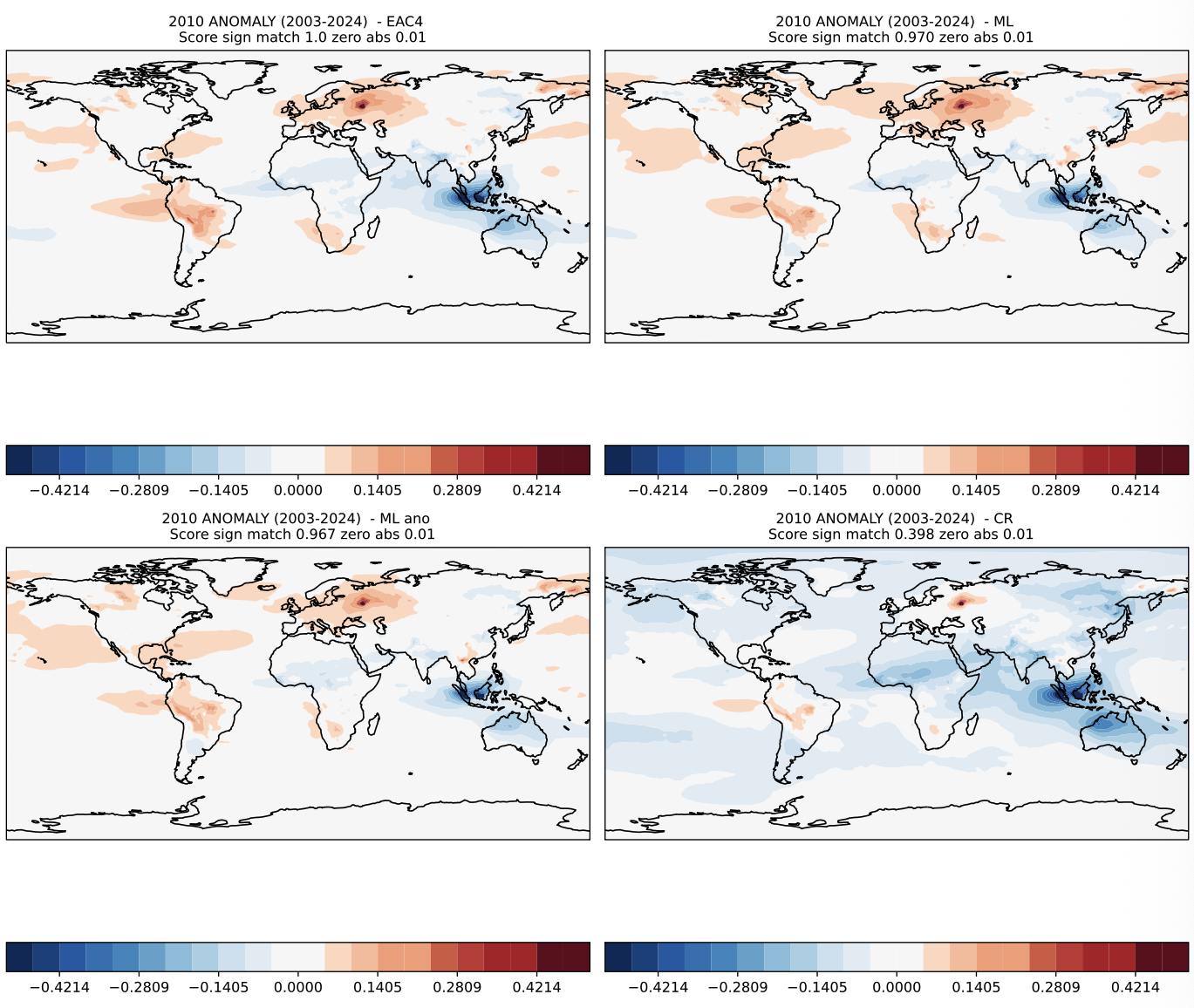}

\caption{.}
\label{monthly_maps}
\end{figure*}

\begin{figure*}[htb]
\includegraphics[width=14cm]{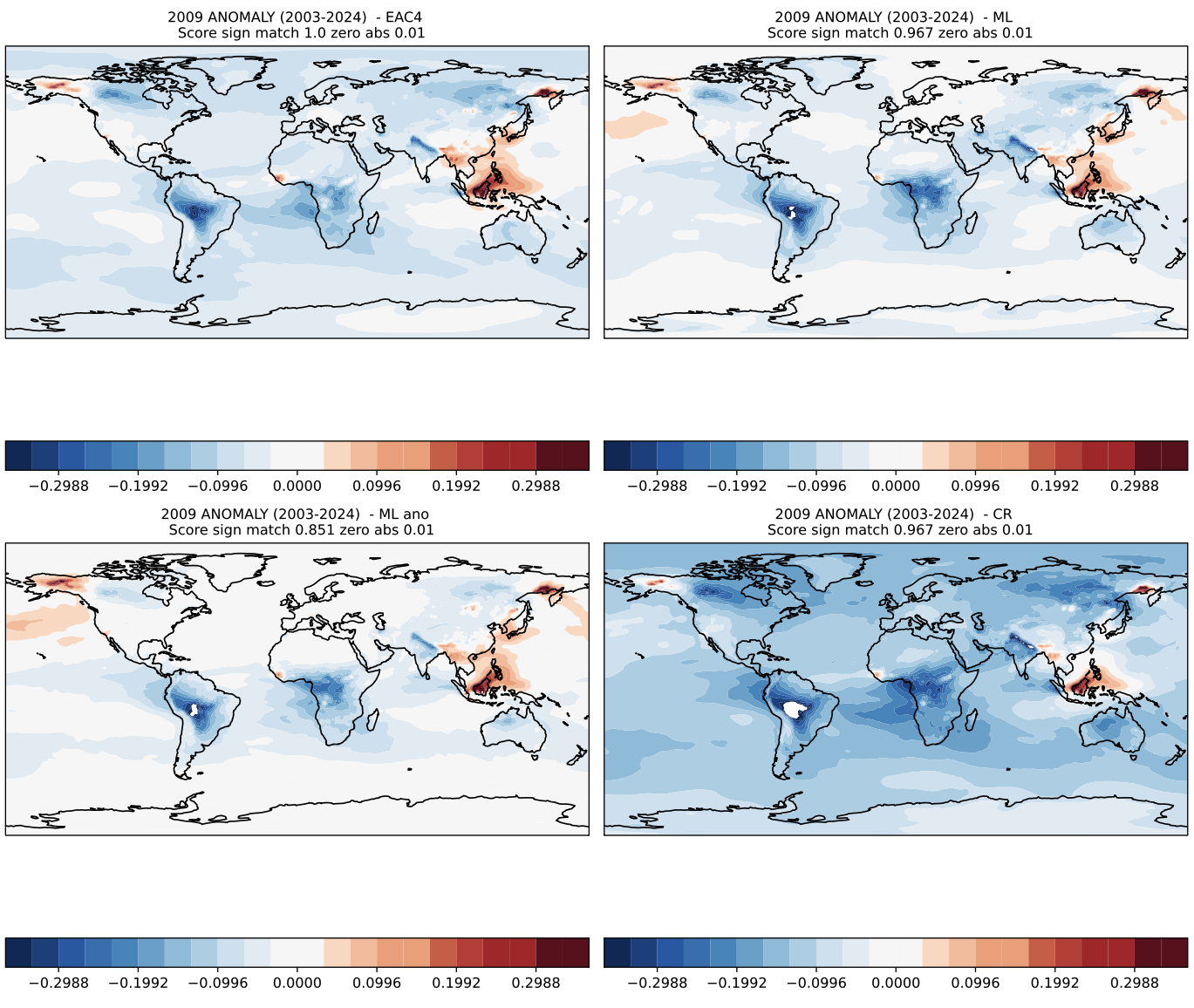}
\includegraphics[width=14cm]{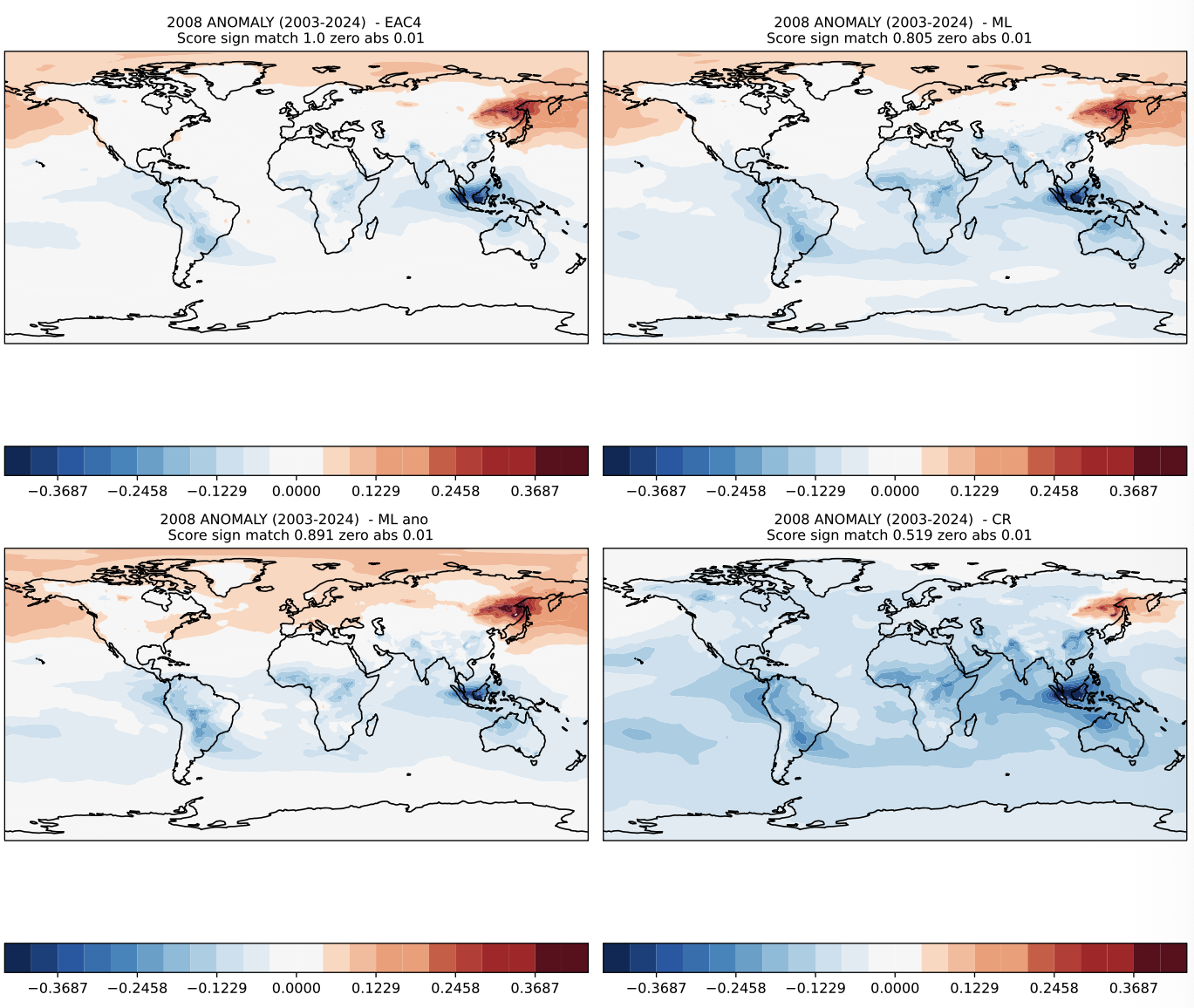}

\caption{.}
\label{monthly_maps}
\end{figure*}

\begin{figure*}[htb]
\includegraphics[width=14cm]{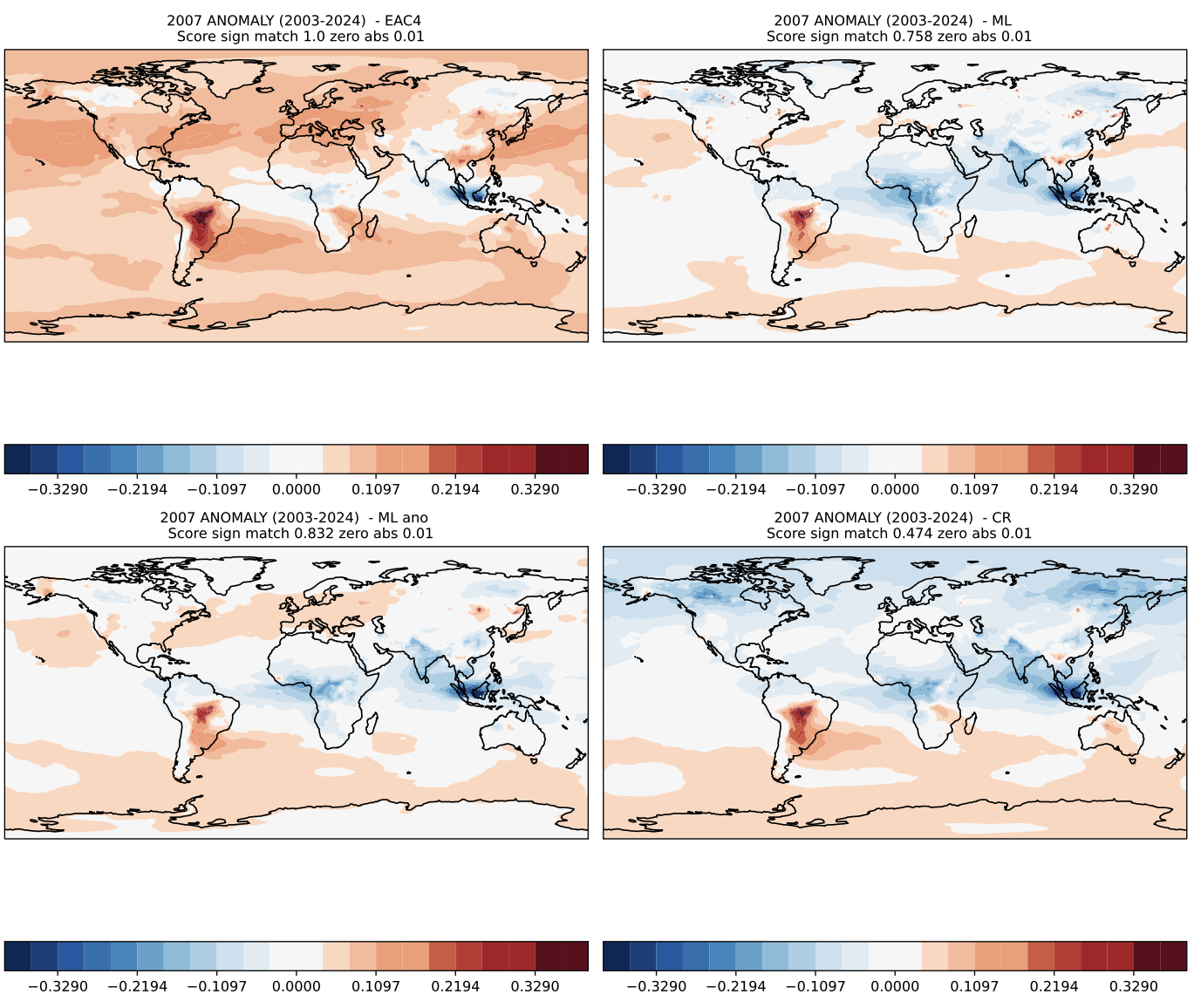}
\includegraphics[width=14cm]{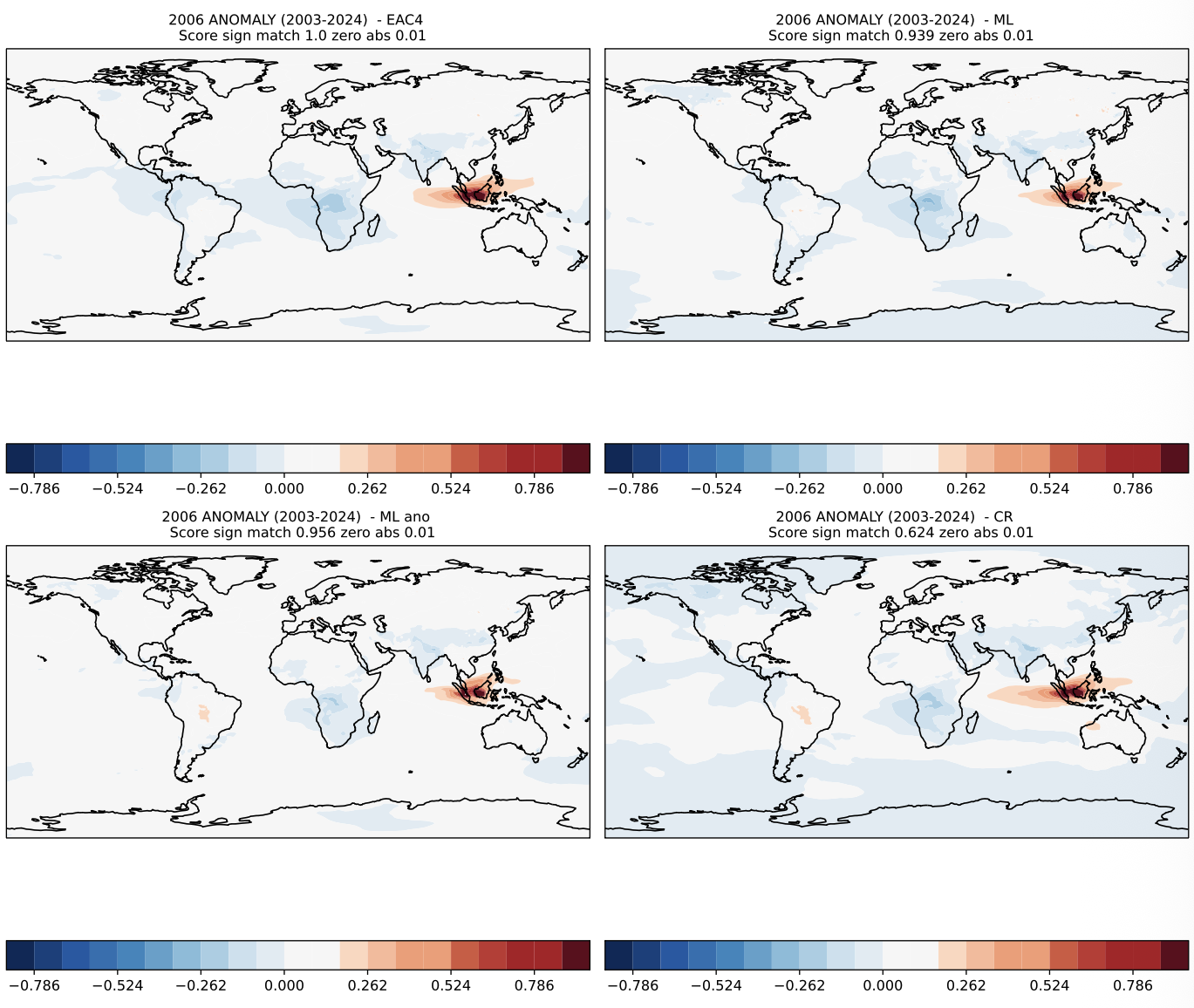}

\caption{.}
\label{monthly_maps}
\end{figure*}

\begin{figure*}[htb]
\includegraphics[width=14cm]{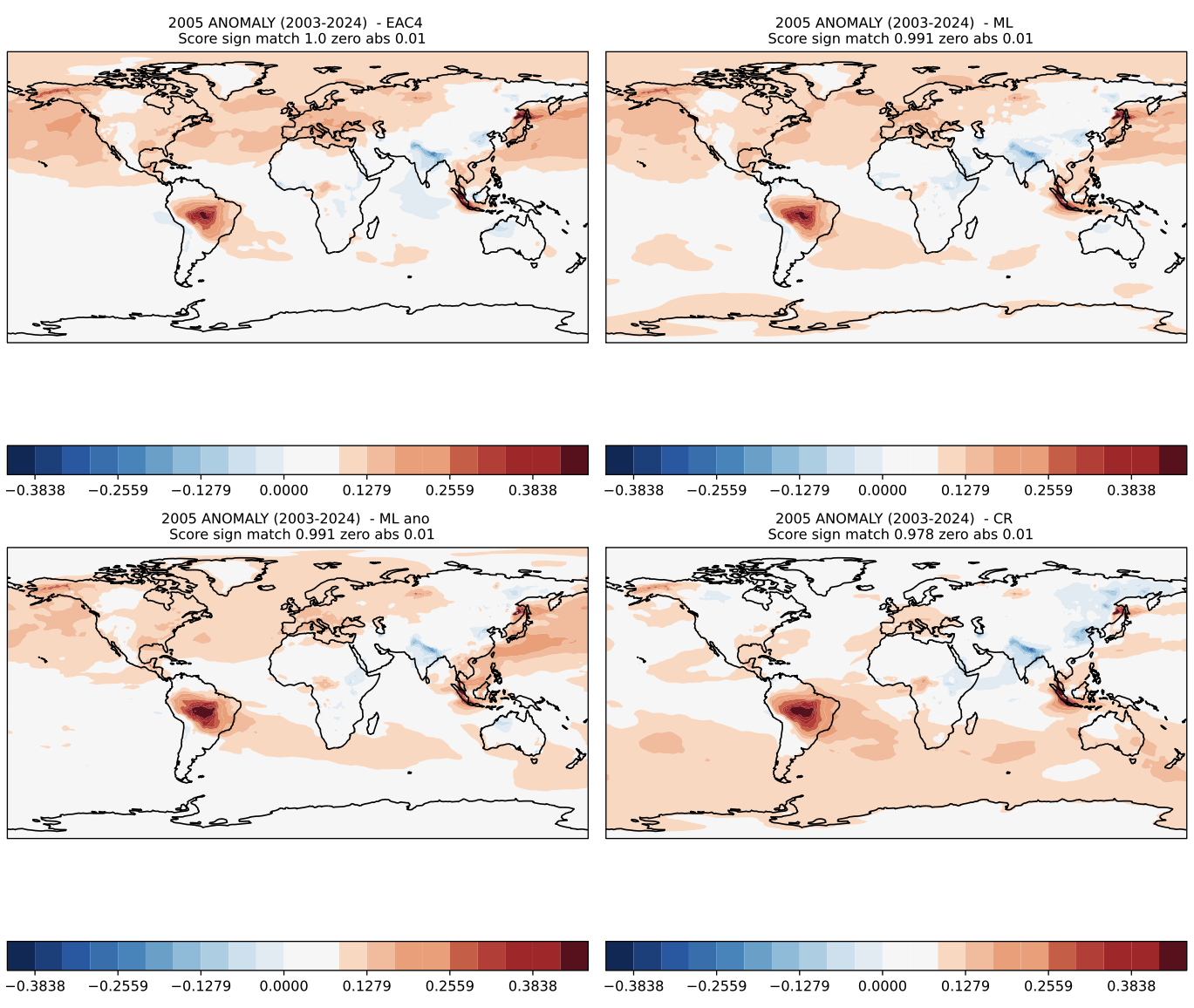}
\caption{Anomaly maps calculate against own mean calculated from 2003-2024.}
\label{monthly_maps}
\end{figure*}

      %% use this to mark the end of the appendix section. Otherwise the figures might be numbered incorrectly (e.g. 10 instead of 1).

%% Regarding figures and tables in appendices, the following two options are possible depending on your general handling of figures and tables in the manuscript environment:

%% Option 1: If you sorted all figures and tables into the sections of the text, please also sort the appendix figures and appendix tables into the respective appendix sections.
%% They will be correctly named automatically.

%% Option 2: If you put all figures after the reference list, please insert appendix tables and figures after the normal tables and figures.
%% To rename them correctly to A1, A2, etc., please add the following commands in front of them:

%\appendixfigures  %% needs to be added in front of appendix figures

%\appendixtables   %% needs to be added in front of appendix tables

%% Please add \clearpage between each table and/or figure. Further guidelines on figures and tables can be found below.

%\authorcontribution{TEXT} %% this section is mandatory

%\competinginterests{TEXT} %% this section is mandatory even if you declare that no competing interests are present

%\disclaimer{TEXT} %% optional section

%\begin{acknowledgements}
%TEXT
%\end{acknowledgements}

%% REFERENCES

%% The reference list is compiled as follows:

\bibliography{mopitt}

\begin{thebibliography}{12}
\providecommand{\natexlab}[1]{#1}
\providecommand{\url}[1]{\texttt{#1}}
\providecommand{\urlprefix}{}
\expandafter\ifx\csname urlstyle\endcsname\relax
  \providecommand{\doi}[1]{https://doi.org/\discretionary{}{}{}#1}\else
  \providecommand{\doi}{https://doi.org/\discretionary{}{}{}\begingroup \urlstyle{rm}\Url}\fi

\bibitem[{Breiman(2001)}]{breiman2001random}
Breiman, L.: Random forests, Machine Learning, 45, 5--32, 2001.

\bibitem[{Deeter et~al.(2003)Deeter, Emmons, Francis, Edwards, Gille, Warner, Khattatov, Ziskin, Lamarque, Ho, Mao, Chen, and Drummond}]{Deeter2003MOPITT}
Deeter, M.~N., Emmons, L.~K., Francis, G.~L., Edwards, D.~P., Gille, J.~C., Warner, J.~X., Khattatov, B., Ziskin, D., Lamarque, J.-F., Ho, S.-P., Mao, D., Chen, J., and Drummond, J.~R.: Operational carbon monoxide retrieval algorithm and selected results for the {MOPITT} instrument, Journal of Geophysical Research: Atmospheres, 108, 4399, \doi{10.1029/2002JD003186}, 2003.

\bibitem[{Flemming et~al.(2015)Flemming, Huijnen, Arteta, Bechtold, Beljaars, Blechschmidt, Diamantakis, Engelen, Gaudel, Inness, Jones, Josse, Katragkou, Marecal, Peuch, Richter, Schultz, Stein, and Tsikerdekis}]{Flemming2015CIFS}
Flemming, J., Huijnen, V., Arteta, J., Bechtold, P., Beljaars, A., Blechschmidt, A.-M., Diamantakis, M., Engelen, R.~J., Gaudel, A., Inness, A., Jones, L., Josse, B., Katragkou, E., Marecal, V., Peuch, V.-H., Richter, A., Schultz, M.~G., Stein, O., and Tsikerdekis, A.: Tropospheric chemistry in the Integrated Forecasting System of ECMWF, Geoscientific Model Development, 8, 975--1003, \doi{10.5194/gmd-8-975-2015}, 2015.

\bibitem[{Friedman(2001)}]{friedman2001greedy}
Friedman, J.~H.: Greedy function approximation: a gradient boosting machine, Annals of statistics, 29, 1189--1232, 2001.

\bibitem[{Inness et~al.(2019)Inness, Ades, Agust{\'i}-Panareda, Barr{\'e}, Benedictow, Blechschmidt, Domingues, Engelen, Eskes, Flemming, Huijnen, Jones, Kipling, Massart, Parrington, Peuch, Razinger, Remy, Schulz, Suttie, and Th{\'e}paut}]{Inness2019EAC4}
Inness, A., Ades, M., Agust{\'i}-Panareda, A., Barr{\'e}, J., Benedictow, A., Blechschmidt, A.-M., Domingues, A., Engelen, R., Eskes, H., Flemming, J., Huijnen, V., Jones, L., Kipling, Z., Massart, S., Parrington, M., Peuch, V.-H., Razinger, M., Remy, S., Schulz, M., Suttie, M., and Th{\'e}paut, J.-N.: The {CAMS} reanalysis of atmospheric composition, Atmospheric Chemistry and Physics, 19, 3515--3556, \doi{10.5194/acp-19-3515-2019}, 2019.

\bibitem[{Kaiser et~al.(2012)Kaiser, Heil, Andreae, Benedetti, Chubarova, Jones, Morcrette, Razinger, Schultz, Suttie, and van~der Werf}]{Kaiser2012GFAS}
Kaiser, J.~W., Heil, A., Andreae, M.~O., Benedetti, A., Chubarova, N., Jones, L., Morcrette, J.-J., Razinger, M., Schultz, M.~G., Suttie, M., and van~der Werf, G.~R.: Biomass burning emissions estimated with a global fire assimilation system based on observed fire radiative power, Biogeosciences, 9, 527--554, \doi{10.5194/bg-9-527-2012}, 2012.

\bibitem[{Kim(2024)}]{kim2024deep_bcsi_pm25}
Kim, e.~a.: Deep-BCSI: A Deep Learning-Based Framework for Bias Correction and Spatial Imputation of PM2.5 Concentrations in South Korea, Atmospheric Research, 301, 107\,283, \doi{10.1016/j.atmosres.2024.107283}, 2024.

\bibitem[{Lary et~al.(2009)Lary, Remer, MacNeill, Roscoe, and Paradise}]{lary2009ml_bias_modis}
Lary, D.~J., Remer, L.~A., MacNeill, D., Roscoe, B., and Paradise, S.: Machine Learning and Bias Correction of MODIS Aerosol Optical Depth, IEEE Geoscience and Remote Sensing Letters, 6, 694--698, \doi{10.1109/LGRS.2009.2023605}, 2009.

\bibitem[{Liu et~al.(2025)Liu, Li, Wild, Doherty, O’Connor, and Turnock}]{liu2025deep_learning_ozone}
Liu, Z., Li, K., Wild, O., Doherty, R.~M., O’Connor, F.~M., and Turnock, S.~T.: Applying Deep Learning to a Chemistry‐Climate Model for Improved Ozone Prediction, Atmospheric Chemistry and Physics, 25, 16\,969--16\,981, \doi{10.5194/acp-25-16969-2025}, 2025.

\bibitem[{Ni et~al.(2026)Ni, Yang, Liao et~al.}]{ni2026ml_bias_ozone}
Ni, Y., Yang, Y., Liao, H., et~al.: Machine Learning‐Based Bias‐Corrected Future Projections of Ozone Concentrations from a Chemistry‐Climate Model, Environmental Science \& Technology, 60, 3135--3147, \doi{10.1021/acs.est.5c11992}, 2026.

\bibitem[{Rumelhart et~al.(1986)Rumelhart, Hinton, and Williams}]{rumelhart1986learning}
Rumelhart, D.~E., Hinton, G.~E., and Williams, R.~J.: Learning representations by back-propagating errors, Nature, 323, 533--536, 1986.

\bibitem[{Zhang et~al.(2024)Zhang, Harrop, Leung, Charalampopoulos, Sorensen, Xu, and Sapsis}]{zhang2024ml_bias_e3sm}
Zhang, S., Harrop, B.~E., Leung, L.~R., Charalampopoulos, A., Sorensen, B.~B., Xu, W., and Sapsis, T.: A Machine Learning Bias Correction of Large‐Scale Environment of High‐Impact Weather Systems in E3SM Atmosphere Model, Journal of Advances in Modeling Earth Systems, 16, \doi{10.1029/2023MS004138}, 2024.

\end{thebibliography}

%% Since the Copernicus LaTeX package includes the BibTeX style file copernicus.bst,
%% authors experienced with BibTeX only have to include the following two lines:
%%
\bibliographystyle{copernicus}

\end{document}